\documentclass[12pt,fleqn]{article}
\usepackage{amsfonts,ulem,url}
\usepackage{amssymb}
\usepackage{amsmath}
\usepackage{amstext,verbatim,graphicx,ifthen,multirow,amsthm}
\usepackage{bm}

\usepackage{natbib}

\renewcommand{\mathbf}{\boldsymbol}
\renewcommand{\thepage}{}
\renewcommand{\appendix}{\footnotesize\parindent 0cm\setcounter{equation}{0}
\renewcommand{\theequation}{A.\arabic{equation}}
\setcounter{lemma}{0}\renewcommand{\thelemma}{A.\arabic{lemma}}}
\newcommand{\oy}{{\overline{Y}}}
\newcommand{\ox}{{\overline{X}}}
\newcommand{\obs}{{\rm obs}}
\newcommand{\mme}{\mathbb{E}}
\newcommand{\bw}{{\mathbf{W}}}

\newcommand{\bww}{{\mathbf{w}}}

\newcommand{\hmucc}{\hat{\mu}_{\rm c}}
\newcommand{\hatx}{\overline{X}_{\rm t}}
\newcommand{\hbetacc}{\hat{\beta}_{\rm c}}

\newcommand{\bxcc}{{\bf X}_{\rm c}}

\newcommand{\xp}{X^\mathrm{p}}
\newcommand{\xr}{X^\mathrm{r}}

\newcommand{\indep}{\perp\!\!\!\perp}

\def\monthname{\ifcase\month\or
  January\or February\or March\or April\or May\or June\or July\or
  August\or September\or October\or November\or December\fi}
\renewcommand{\thepage}{[\arabic{page}]}
\numberwithin{equation}{section}

\def\monthname{\ifcase\month\or
January\or February\or March\or April\or May\or June\or
July\or August\or September\or October\or November\or December\fi}
\renewcommand{\appendix}{\small\parindent 0cm\setcounter{equation}{0}
\renewcommand{\theequation}{A.\arabic{equation}}
\setcounter{lemma}{0}\renewcommand{\thelemma}{A.\arabic{lemma}}
\setcounter{theorem}{0}\renewcommand{\thetheorem}{A.\arabic{theorem}}}
\begin{document}

\title{The State of Applied Econometrics - Causality and Policy Evaluation
\thanks{{\small We are
grateful for comments .}} }
\author{Susan Athey\thanks{{\small Graduate School of Business, Stanford University, and NBER,
athey@stanford.edu. }} \and
Guido W. Imbens\thanks{{\small 
Graduate School of Business, Stanford University, and NBER,
imbens@stanford.edu. }}
 }
\date{ 
\ifcase\month\or
January\or February\or March\or April\or May\or June\or
July\or August\or September\or October\or November\or December\fi \ \number%
\year\ \  }
\maketitle

\begin{abstract}
In this paper we discuss recent developments in econometrics that we view as important for empirical researchers working on policy evaluation questions. We focus on three main areas, where in each case we highlight recommendations for applied work. First, we discuss new research on identification strategies in program evaluation, with particular focus on synthetic control methods, regression discontinuity,  external validity, and the causal interpretation of regression methods. Second, we discuss various forms of   supplementary analyses to make the identification strategies more credible. These include placebo analyses as well as sensitivity and robustness analyses.
Third, we discuss recent advances in machine learning methods for causal effects. These advances include methods to adjust for differences between treated and control units in high-dimensional settings, and methods for identifying and estimating heterogenous treatment effects.
\end{abstract}

\begin{center}
\end{center}


\pagestyle{plain}
\baselineskip=20pt\newpage \setcounter{page}{1}\renewcommand{\thepage}{[%
\arabic{page}]}\renewcommand{\theequation}{\arabic{section}.%
\arabic{equation}}

\section{Introduction}
\label{section:introduction}


This article synthesizes recent developments in econometrics that may be useful for researchers interested in estimating the effect of policies on outcomes.  For example, what is the effect of the minimum wage on
employment? Does improving educational outcomes for some students spill over onto other students? Can we credibly estimate the effect of labor market interventions with observational studies? Who benefits from job training programs?  We focus on the case where the policies of interest had been implemented for 
at least some units in an available dataset, and the outcome of interest is also observed in that dataset.  We do not consider here questions about outcomes that cannot be directly
measured in a given dataset, such as consumer
welfare or worker well-being, and we do not consider questions about policies that have never been implemented.  The latter type of question is considered in a branch of applied work
referred to as ``structural'' analysis; the type of analysis considered in this review is sometimes referred to as ``reduced-form,'' or ``design-based,'' or ``causal'' methods.

The gold standard for drawing inferences about the effect of a policy is the randomized controlled experiment; with data from a randomized experiment, by construction those units who were exposed
to the policy are the same, in expectation, as those who were not, and it becomes relatively straightforward to draw inferences about the causal effect of a policy.  The difference between the sample average outcome for treated units and control units is an unbiased estimate of the average causal effect.  Although digitization has lowered the costs of conducting randomized experiments in many settings, it remains the case that many policies are expensive to test experimentally.  In other
cases, large-scale experiments may not be politically feasible.  For example, it would be challenging to randomly allocate the level of minimum wages to different states or metropolitan areas in the
United States.  Despite the lack of such randomized controlled experiments, policy makers still need to make decisions about the minimum wage.  A large share of the empirical work
in economics about policy questions relies on observational data--that is, data where policies were determined in a way other than random assignment. But drawing inferences about the causal effect of a policy from observational data is quite challenging.

To understand the challenges, consider the example of the minimum wage.  It might be the case that states with higher costs of living, as well as more price-insensitive consumers,
select higher levels of the minimum wage.  Such states might also see employers pass on higher wage costs to consumers without losing much business.  In contrast, states with lower cost
of living and more price-sensitive consumers might choose a lower level of the minimum wage.  A naive analysis of the effect of a higher minimum wage on employment might compare
the average employment level of states with a high minimum wage to that of states with a low minimum wage.  This difference is {\it not} a credible estimate of the causal effect of a higher minimum wage: it is not a good estimate of
the change in employment that would occur if the low-wage state raised its minimum wage.  The naive estimate would confuse correlation with causality.  In contrast, if the minimum
wages had been assigned randomly, the average difference between low-minimum-wage states and high-minimum-wage states would have a causal interpretation.

Most of the attention in the econometrics literature on reduced-form policy evaluation focuses on issues surrounding separating correlation from causality in observational studies, that is, with non-experimental data.  There are several distinct
strategies for estimating causal effects with observational data.  These strategies are often referred to as ``identification strategies,'' or ``empirical strategies'' (\citet{angristkrueger_strategies}) because they are strategies for ``identifying'' the causal effect.  We say that a causal effect is ``identified'' if it can be learned when the data set is sufficiently large.
  Issues of identification are distinct from issues that arise because of limited data. 
 In Section \ref{section:evaluation}, we review recent developments corresponding to several different
identification strategies.  An example of an identification strategy is one based on ``regression discontinuity.'' This type of strategy can be used in a setting when allocation to a treatment is based on a ``forcing'' variable, such as location, time, or
birthdate being above or below a threshold.  For example, a birthdate cutoff may be used for school entrance or for the decision of whether a child can legally drop out of school in a given academic year; and there may be geographic boundaries for assigning students to schools or patients to hospitals.  
The identifying assumption is that there is no discrete change in the characteristics of individuals who fall on one side
or the other of the threshold for treatment assignment.  Under that assumption, the relationship between outcomes and the forcing variable can be modeled, and deviations from the predicted relationship at the treatment assignment boundary can be attributed to the treatment.
Section  \ref{section:evaluation} also considers other strategies such as synthetic control methods, methods designed for networks settings, and methods that combine experimental and observational data.

In Section \ref{section:suppl} we discuss what we refer to in general as {\it supplementary analyses}. By supplementary analyses we mean analyses where the focus is on providing support for the identification strategy underlying the primary analyses, on establishing
that the modeling decisions are adequate to capture the critical features of the identification strategy, or on establishing robustness of estimates to modeling assumptions. Thus the results of the supplementary analyses are intended to convince the reader of the credibility of the primary analyses. Although these analyses often involve  statistical tests, the focus is not on goodness of fit measures. Supplementary analyses can take on a variety of forms, and we discuss some of the most interesting ones that have been proposed thus far. In our view these supplementary analyses will be of growing importance for empirical researchers.  In this review, our goal is to organize these analyses,
which may appear to be applied unsystematically in the empirical literature, or may have not received a lot of formal attention in the econometrics literature.

In Section \ref{section:machine} we  discuss briefly new developments coming from what is referred to as the machine learning literature.  Recently there has been much interesting work combining these predictive methods with causal analyses, and this is the part of the literature that we put special emphasis on in our discussion.  We show how machine learning methods can be used to deal with datasets with 
many covariates, and how they can be used to enable the researcher to build more flexible models.  Because many common 
identification strategies rest on assumptions such as the ability of the researcher to observe and control for confounding variables
(e.g. the factors that affect treatment assignment as well as outcomes), or to flexibly model the factors that affect outcomes in the 
absence of the treatment, machine learning methods hold great promise in terms of improving the credibility of policy evaluation, and 
they can also be used to approach supplementary analyses more systematically.

As the title indicates, this review is limited to methods relevant for policy analysis, that is, methods for causal effects. Because  there is another review in this issue focusing on structural methods, as well as one on theoretical econometrics, we largely  refrain from discussing those areas, focusing more narrowly on what is sometimes referred to as reduced-form methods, although we prefer the terms causal or design-based methods, with an emphasis on recommendations for applied work. 
The choices for topics within this area is based on our reading of recent research, including ongoing work, and we point out areas where we feel there are interesting open research questions. This is of course a subjective perspective.

\section{New Developments in Program Evaluation}
\label{section:evaluation}

The econometric literature on estimating causal effects has been a very active one for over three decades now. 
Since the early 1990s the potential outcome, or Neyman-Rubin Causal Model, approach to these problems has gained substantial acceptance as a framework for analyzing causal problems. (We should note, however, that there is a complementary approach based on graphical models (e.g., \citet{pearl}) that is widely used in other disciplines, though less so in economics.) 
In the potential outcome approach, there is for each unit $i$, and  each level of the treatment $w$, a potential outcome $Y_i(w)$, that describes the level of the outcome under treatment level $w$ for that unit. In this perspective, causal effects are comparisons of pairs of potential outcomes for the same unit, e.g., the difference $Y_i(w')-Y_i(w)$. Because a given unit can only receive one level of the treatment, say $W_i$, and only the corresponding level of the outcome, $Y_i^\obs=Y_i(W_i)$ can be observed, we can never directly observe the causal effects, which is what \citet{holland} calls the ``fundamental problem of causal inference.'' Estimates of causal effects are ultimately based on comparisons of different units with different levels of the treatment.

A large part of the causal or treatment effect literature has focused on estimating average treatment effects in a binary treatment setting under the unconfoundedness assumption (e.g., \citet{rosenbaum1983central}),
\[ W_i\ \indep\ \Bigl(Y_i(0),Y_i(1)\Bigr)\ \Big|\ X_i.\]
 Under this assumption, associational or correlational relations such as $\mathbb{E}[Y^\obs_i|W_i=1,X_i=x]-\mathbb{E}[Y^\obs_i|W_i=0,X_i=x]$ can be given a causal interpretation as the average treatment effect $\mathbb{E}[Y_i(1)-Y_i(0)|X_i=x]$. The literature on estimating average treatment effects under unconfoundedness is by now a very mature literature, with a number of competing estimators and many applications. Some estimators use matching methods, some rely on weighting, and some involve the propensity score, the conditional probability of receiving the treatment given the covariates, $e(x)={\rm pr}(W_i=1|X_i=x)$. There are a number of recent reviews of the general literature (\citet{imbens2004,  imbens2015causal}, and for a different perspective \citet{heckman, heckman2007}), and we do not review it in its entirety in this review. However, one area with continuing developments concerns settings  with many covariates, possibly more than there are units. For this setting connections have been made with the machine learning and big data literatures. We review these new developments in  Section \ref{subsection:machine_average}. 
In the context of many covariates there has also been interesting developments in estimating heterogenous treatment effects;  we cover this literature in Section \ref{subsection:machine_hetero}.
 We also discuss, in Section \ref{subsection:multi}, settings with unconfoundedness and multiple levels for the treatment.

Beyond settings with unconfoundedness we discuss issues related to a number of other identification strategies and settings.
In Section \ref{subsection:rdd},
 we  discuss regression discontinuity designs.
Next, we discuss synthetic control methods as developed in  the \citet{abadie2010},  which we believe is one the most important development in program evaluation in the last decade.
 In Section \ref{subsection:networks} we discuss causal methods in network settings. In Section \ref{subsection:regression} we  draw attention to some recent work on the causal interpretation of regression methods.  We also discuss external validity in Section \ref{subsection:external}, and finally, in Section 
\ref{subsection:leveraging} we discuss how randomized experiments can provide leverage for observational studies.

In this review we 
  do not discuss  the recent literature on instrumental variables. There are two major  strands of that by now fairly mature literature. One focuses on heterogenous treatment effects, with a key development the notion of the local average treatment effect (\citet{imbens1994, angrist1996}). This literature has recently been reviewed in \citet{imbens2014}. There is also a separate literature on weak instruments, focusing on settings with a possibly large number of instruments and weak correlation between the instruments and the endogenous regressor. See \citet{bekker1994, stock1997, chamberlain} for specific contributions, and \citet{andrews2007} for a survey. We also do not discuss in detail bounds and partial identification analyses. Since the work by Manski (e.g., \citet{manski_bounds}) these have received a lot of interest, with an excellent recent review in \citet{tamer}.

\subsection{Regression Discontinuity Designs}
\label{subsection:rdd}

A regression discontinuity design is a research design that exploits discontinuities in incentives to participate in a treatment to evaluate the effect of these treatment.

\subsubsection{Set Up}

In regression discontinuity designs, we are interested in the causal effect of a binary treatment or program, denoted by $W_i$. The key feature of the design is the presence of an exogenous variable, the forcing variable, denoted by $X_i$, such that at a particular value of this forcing variable, the threshold $c$, the probability of participating in the program or being exposed to the treatment changes discontinuously:
\[ \lim_{x\uparrow c} {\rm pr}(W_i=1|X_i=x)\neq 
\lim_{x\downarrow c} {\rm pr}(W_i=1|X_i=x).\]
If the jump in the conditional probability is from zero to one, we have a {\it sharp} regression discontinuity (SRD) design; if the magnitude of the jump is less than one, we have a {\it fuzzy} regression discontinuity  (FRD) design.
The estimand is the discontinuity in the conditional expectation of the outcome at the threshold, scaled by the discontinuity in the probability of receiving the treatment:
\[ \tau^{\rm rd}=\frac{\lim_{x\downarrow c}\mme[Y_i|X_i=x]-
\lim_{x\uparrow c}\mme[Y_i|X_i=x]}{\lim_{x\downarrow c}\mme[W_i|X_i=x]-\lim_{x\uparrow c}\mme[W_i|X_i=x]}.\]
In the SRD case the denominator is equal to one, and we just focus on the discontinuity of the conditional expectation of the outcome given the forcing variable at the threshold.
In that case, under the assumption that the individuals just to the right and just to the left of the threshold are comparable, the estimand has an interpretation as the average effect of the treatment for individuals close to the threshold. In the FRD case, the interpretation of the estimand is the average effect for compliers at the threshold (i.e., individuals at the threshold whose treatment status would have changed had they been on the other side of the threshold) \citep{hahntodd}.

\subsubsection{Estimation and Inference}

In the general FRD case, the estimand $\tau^{\rm rd}$ has four components, each of them  the limit of the conditional expectation of a variable at a particular value of the forcing variable. We can think of this, after splitting the sample by whether the value of the forcing variable exceeds the threshold or not, as estimating the conditional expectation at a boundary point.
Researchers typically wish to use flexible (e.g., semiparametric or nonparametric) methods for estimating these conditional expectations. Because the target in each case is the conditional expectation at a boundary point, simply differencing average outcomes close to the threshold on the right and on the left leads to an estimator with poor properties, as stressed by \citet{porter}. As an alternative \citet{porter} suggested ``local linear regression,'' which involves estimating linear regressions of outcomes on the forcing variable separately on the left and the right of the 
threshhold,
weighting most heavily observations close to the threshold, and then taking the difference between the predicted values at the threshold.
This local linear estimator has substantially better finite sample properties than nonparametric methods that do not account for threshold effects, and it has become the standard.  There are some suggestions that using local quadratic methods may work well given the current technology for choosing bandwidths (e.g., \citet{calonico}). Some applications use global high order polynomial approximations to the regression function, but there has been some criticism of this practice. \citet{gelmanimbens} argue that in practice it is difficult to choose the order of the polynomials in a satisfactory way, and that confidence intervals based on such methods have poor properties.

Given a local linear estimation method, a key issue  is the choice of the bandwidth, that is, how close observations need to be to the threshold. Conventional methods for choosing optimal bandwidths in nonparametric estimation, e.g., based on cross-validation, look for bandwidths that are optimal for estimating the entire regression function, whereas here the interest is solely in the value of the regression function at a particular point. 
The current state of the literature suggests choosing the bandwidth for the local linear regression using asymptotic expansions of the estimators around small values for the bandwidth. See \citet{imbenskalyanaraman} and \citet{cattaneo2010} for further discussion.

In some cases, the discontinuity involves multiple exogenous variables. For example, in  \citet{jacob} and \citet{matsudaira}, the focus is on the causal effect of attending summer school. The formal rule is that students who score below a threshold on either a language or a mathematics test are required to attend summer school. Although not all the students who are required to attend summer school do so (so that this a fuzzy regression discontinuity design), the fact that the forcing variable is a known function of two observed exogenous variables makes it possible to estimate  the effect of summer school at different margins. For example, one can estimate of the effect of summer school for  individuals who are required to attend summer school because of failure to pass the language test, and compare this with the estimate for those who are required because of failure to pass the mathematics test. Even more than the presence of other exogenous variables, the dependence of the threshold on multiple exogenous variables improves the ability to detect and analyze heterogeneity in the causal effects.

\subsubsection{An Illustration}

Let us illustrate the  regression discontinuity design with data from \citet{jacob}. 
 \citet{jacob} use administrative data from the Chicago Public Schools which instituted in
1996 an accountability policy that tied summer school attendance and promotional decisions to performance on standardized tests.
We use the data for 70,831 third graders in years 1997-99.
The rule was that individuals score below the threshold (2.75 in this case)  on either a reading or mathematics score before the summer were required to attend summer school. 
It should be noted that the initial scores range from 0 to 6.8, with increments equal to 0.1.
The outcome variable $Y_i^{\rm obs}$ is  the math score after the summer school, normalized to have variance one.
Out of the 70,831 third graders,        15,846 score below  the threshold on the mathematics test,
       26,833 scored below the threshold on the reading test,       12,779 score below the threshold on both tests, and        29,900 scored below the threshold on at least one test.

Table \ref{tabel1} presents some of the results. The first  row presents an estimate of the effect on the mathematics test, using  for the forcing variable
the minimum of the initial mathematics score and the initial reading score. 
We find that the program has a substantial effect.
Figure 1 shows which students contribute to this estimate. 
The figure shows a scatterplot of 1.5\% of the students, with uniform noise added to their actual scores to show the distribution more clearly. The solid line shows the set of values for the mathematics and reading scores that would require the students to participate in the summer program. The area enclosed by the dashed line contains all the students within the bandwidth from the threshold.

We can partition the sample into students with relatively high reading scores (above the threshold plus the Imbens-Kalyanaraman bandwidth), who could only be in the summer program because of their mathematics score, students with relatively high mathematics scores (above the threshold plus the bandwidth) who could only be in the summer program because of their reading score, and students with low mathematics and reading scores (below the threshold plus the bandwidth).
 Rows 2-4 present estimates for these separate subsamples. We find that there is relatively little evidence of heterogeneity in the estimates of the program.

The last row demonstrates the importance of using local linear rather than standard kernel (local constant) regressions. Using the same bandwidth, but using a weighted average of the outcomes rather than a weighted linear regression, leads to an estimate equal to -0.15: rather than benefiting from the summer school, this estimate counterintuitively suggests that the summer program hurts the students in terms of subsequent performance. This bias that leads to these negative estimates  is not surprising: the students who participate in the program are on average worse  in terms of prior performance than the students who do not participate in the program, even if we only use information for students close to the threshold.

 \begin{table}[ht]
 \caption{\sc  Regression Discontinuity Designs: The Jacob-Lefgren Data}
 \vskip1cm
 \begin{center}
 \begin{tabular}{lllccc}
 \\
Outcome & Sample& Estimator& Estimate & (s.e.) & IK Bandwidth\\
\\\hline
 Math  &  All& Local Linear&0.18 & (0.02) & 0.57\\
Math  & Reading $>3.32$  &Local  Linear &0.15 & (0.02) & 0.57\\
Math  & Math $>3.32$& Local Linear&0.17 & (0.03) & 0.57\\
Math  & Math and Reading $<3.32$& Local Linear&0.19 & (0.02) & 0.57\\
\\
 Math  & All&Local Constant &-0.15 & (0.02) & 0.57\\
 \end{tabular}
 \end{center}
 \label{tabel1}
\end{table}

\subsubsection{Regression Kink Designs}

One of the most interesting recent developments in the area of regression discontinuity designs is the generalization to discontinuities in derivatives, rather than levels, of conditional expectations. The first discussions of these regression kink designs are in
\citet{nielsen,cardkink,dong2}. The basic idea is that at a threshold for the forcing variable, the slope of the outcome function (as a function of the forcing variable) changes, and the
goal is to estimate this change in slope.   

To make this clearer, let us discuss the example in \citet{cardkink}. The forcing variable is a lagged earnings variable that determines unemployment benefits. A simple rule would be that unemployment benefits are a fixed percentage of last year's earnings, up to a maximum. Thus the unemployment benefit, as a function of the forcing variable, is a continuous, piecewise linear function. Now suppose we are interested in the causal effect of an increase in the unemployment benefits on the duration of unenmployment spells. Because the benefits are a deterministic function of lagged earnings, direct comparisons of individuals with different levels of benefits are confounded by differences in lagged earnings. However, at the threshold, the relation between benefits and lagged earnings changes. Specifically, the derivative of the benefits with respect to lagged earnings changes. If we are willing to assume that in the absence of the kink in the benefit system, the derivative of the expected duration would be smooth in lagged earnings, then the change in the derivative of the expected duration with respect to lagged earnings is informative about the relation between the expected duration and the benefit schedule, similar to the identification in a regular regression discontinuity design.

To be more precise, suppose the benefits as a function of lagged earnings satisfy
\[ B_i=b(X_i),\]
with $b(x)$ known and continuous, with a discontinuity in the first derivative at $x=c$. Let $b'(v)$ denote the derivative, letting $b'(c+)$ and $b'(c-)$ denote the derivatives from the right and the left at $x=c$. If the benefit schedule is piecewise linear, we would have
\[ B_i=
\beta_{0}+\beta_{1-}\cdot (X_i-c),\ \ X_i<c,\]
\[ B_i=\beta_{0}+\beta_{1+}\cdot (X_i-c),\ \ X_i\geq c.\]
This relationship is deterministic, making this a sharp regression kink design.
Here, as before, $c$ is the threshold. The forcing variable $X_i$ is lagged earnings, $B_i$ is the unemployment benefit that an individual would receive. As  a function of the benefits $b$, the logarithm of the unemployment duration, denoted by $Y_i$, is assumed to satisfy
\[Y_i(b)=\alpha+\tau\cdot \ln (b)+\varepsilon_i.\]
Let $g(x)=\mme[Y_i|X_i=x]$ be the conditional expectation of $Y_i$ given $X_i=x$, with derivative $g'(x)$. The derivative is assumed to exist everywhere other than at $x=c$, where the limits from the right and the left exist.
The idea is to characterize $\tau$ as
\[ \tau=\frac{\lim_{x\downarrow c} g'(x)-\lim_{x\uparrow c} g'(x)}{\lim_{x\downarrow c} b'(x)-\lim_{x\uparrow c} b'(x)}.\]
\citet{cardkink} propose estimating $\tau$ by first estimating $g(x)$ by local linear or local quadratic regression around the threshold. We then divide the difference in the estimated derivative from the right and the left by the difference in the derivatives of $b(x)$ from the right and the left at the threshold.

In some cases, the relationship between $B_i$ and $X_i$ is not deterministic, making it a fuzzy regression kink design. In the fuzzy version of the regression kink design, the conditional expectation of $B_i$ given $X_i$ is estimated using the same approach to get an estimate of the change in the derivative at the threshold.

\subsubsection{Summary of Recommendations}

There are some specific choices to be made in regression discontinuity analyses, and here we provide our recommendations for these choices.
We recommend using local linear or local quadratic methods (see for details on the implementation \citet{hahntodd, porter, calonico}) rather than global polynomial methods.  \citet{gelmanimbens} present a detailed discussion on the concerns with global polynomial methods. These local linear methods require a bandwidth choice. We recommend the optimal bandwidth algorithms based on asymptotic arguments involving local expansions  discssed in \citet{imbenskalyanaraman, calonico}. We also recommend carrying out supplementary analyses to assess the credibility of the design, and in particular to test for evidence of manipulation of the forcing variable. Most important here is the McCrary   test for discontinuities in the density of the forcing variable (\citet{mccrary}), as well as tests for discontinuities in average covariate values at the threshold. We discuss examples of these  in the section on supplementary analyses (Section \ref{subsection:suppl_rdd}). We also recommend researchers to investigate external validity of the regression discontinuity estimates by assessing the credibility of extrapolations to other subpopulations (\citet{bertanhaimbens, angristrokkanen, angristfernandez, donglewbel}). See Section \ref{subsection:external} for more details.

\subsubsection{The Literature}
Regression Discontinuity Designs have a long history, going back to work in psychology in the fifties by \citet{campbell}, but the methods did not become part of the mainstream economics literature until the early 2000s (with \citet{goldberger1, goldberger2} an exception). Early applications in economics include \citet{black} \citet{angrist1999}, 
\citet{vanderklaauw}, \citet{lee}. Recent reviews include \citet{imbenslemieux, leelemieux, vanderklaauw2, titiunik}. More recently there have been many applications (e.g., \citet{greenstone}) and a substantial amount of new theoretical work which has led to substantial improvements in our understanding of these methods. 

\subsection{Synthetic Control Methods and Difference-In-Differences}
\label{subsection:synthetic}

Difference-In-Differences (DID) methods have become an important tool for empirical researchers.
In the basic setting there are two or more groups, at least one treated and one control, and we observe (possibly different) units from all groups in two or more time periods, some prior to the treatment and some after the treatment. The difference between the treatment and control groups post treatment is adjusted for the difference between the two groups prior to the treatment. In the simple DID case these adjustments are linear: they take the form of estimating the average treatment effect as the difference in average outcomes post treatment minus the difference in average outcomes pre treatment.
 Here we discuss two important recent developments, the synthetic control approach and the nonlinear changes-in-changes method.

\subsubsection{Synthetic Control Methods}

Arguably the most important innovation in the evalulation literature in the last fifteen years is the synthetic control approach developed by \citet{abadie2010,abadie2014} and \citet{abadie2003}. This method  builds on  difference-in-differences estimation, but uses arguably more attractive comparisons to get causal effects.
We discuss the basic \citet{abadie2010} approach, and highlight  alternative choices and restrictions that may be imposed to further improve the performance of the methods relative to difference-in-differences estimation methods.

 We observe outcomes for a number of units, indexed by $i=0,\ldots,N$, for a number of periods indexed by $t=1,\ldots,T$.
There is a single unit, say unit $0$, who was exposed to the control treatment during periods $1,\ldots,T_0$ and who received the active treatment, starting in period $T_0+1$. 
For ease of exposition let us focus on the case with $T=T_0+1$ so there is only a single post-treatment period.
All other units are exposed to the control treatment for all periods. 
The number of control units $N$ can be as small as 1, and the number of periods $T$ can be as small as 2.
We may also observe exogenous fixed covariates for each of the units.
The units are often aggregates of individuals, say states, or cities, or countries.
 We are interested in the causal effect of the treatment for this unit, $Y_{0T}(1)-Y_{0T}(0)$.

The traditional DID approach would compare the change for the treated unit (unit 0) between periods $t$ and $T$, for some $t<T$, to the corresponding change for some other unit. For example, consider the classic difference-in-differences study by \citet{cardmariel}. Card is interested in the effect of the Mariel boatlift, which brought Cubans to Miami, on the Miami labor market, and specifically on the wages of low-skilled workers. He compares the change in the outcome of interest, for Miami, to the corresponding change in a control city. He considers various possible control cities, including Houston, Petersburg, Atlanta.

The synthetic control idea is to move away from using a single control unit or a simple average of control units, and instead use a weighted average of the set of controls, with the weights chosen so that the weighted average is similar to the treated unit in terms of lagged outcomes and covariates. In other words, instead of choosing between  Houston, Petersburg or Atlanta, or taking a simple average of outcomes in those cities, the synthetic control approach chooses weights $\lambda_{\rm h}$, $\lambda_{\rm p}$, and $\lambda_{\rm a}$ for Houston, Petersburg and Atlanta respectively, so that
$\lambda_{\rm h}\cdot Y_{{\rm h}t}+
\lambda_{\rm p}\cdot Y_{{\rm p}t}+
\lambda_{\rm a}\cdot Y_{{\rm a}t}$ is close to $Y_{{\rm m}t}$ (for Miami) for the pre-treatment periods $t=1,\ldots,T_0$, as well as for the other  pretreatment variables  (e.g., \citet{peri2015}). This is a very simple, but very useful idea. Of course, if pre-boatlift wages are higher in Houston than in Miami, and higher in Miami than in Atlanta, it would make sense to compare Miami to the average of Houston and Atlanta rather than to Houston  or Atlanta. The simplicity of the idea, and the obvious improvement over the standard methods, have made this a widely used method in the short period of time since its inception.

The implementation of the synthetic control method requires a particular choice for estimating the weights. The original paper \citet{abadie2010} restricts the weights to be non-negative and requires them to add up to one. Let $K$ be the dimension of the covariates $X_i$, and let $\Omega$ be an arbitrary positive definite $K\times K$ matrix. Then let $\lambda(\Omega)$ be the weights that solve
\[ \lambda(\Omega)=\arg\min_\lambda \left(X_0-\sum_{i=1}^N \lambda_i\cdot X_{i}\right)'
\Omega
\left(X_0-\sum_{i=1}^N \lambda_i\cdot X_{i}\right).\]
 \citet{abadie2010} choose the weight matrix $\Omega$ that minimizes
\[ \sum_{t=1}^{T_0} 
\left(Y_{0t}-\sum_{i=1}^N \lambda_i(\Omega)\cdot Y_{it}\right)^2.\]
If the covariates $X_i$ consist of the vector of lagged outcomes, this estimate amounts to minimizing
\[ \sum_{t=1}^{T_0} 
\left(Y_{0t}-\sum_{i=1}^N \lambda_i\cdot Y_{it}\right)^2,\]
subject to the restrictions that the $\lambda_i$ are non-negative and summ up to one.

\citet{doudchenko} point out that one can view the question of estimating the weights in the Abadie-Diamond-Hainmueller synthetic control method differently. Starting with the case without covariates and only lagged outcomes, one can consider the regression function
\[ Y_{0t}=\sum_{i=1}^N \lambda_i\cdot Y_{it}+\varepsilon_t,\]
with $T_0$ units and $N$ regressors.
The absence of the covariates is rarely important, as the fit typically is driven by  matching up the lagged outcomes rather than matching the covariates.
Estimating this regression by least squares is typically not possible because the number of regressors $N$ (the number of control units) is often larger than, or the same order of magnitude as, the number of observations (the number of time periods $T_0$). We therefore need to regularize the estimates in some fashion or another. There are a couple of natural ways to do this. \citet{abadie2010} impose the restriction that the weights $\lambda_i$ are non-negative and add up to one. That often leads to a unique set of weights. However, there are alternative ways to regularize the estimates. In fact, both the restrictions that \citet{abadie2010} impose may hurt performance of the model. If the unit is on the extreme end of the distribution of units, allowing for weights that sum up to a number different from one, or allowing for negative weights may improve the fit. We can do so by using alternative regularization methods such as best subset regression, or LASSO (see Section \ref{subsubsection:LASSO} for a description of LASSO) where we add a penalty proportional to the sum of the weights.
\citet{doudchenko} explore such approaches.


\subsubsection{Nonlinear Difference-in-Difference Models}

A commonly noted concern with difference-in-difference methods is that functional form assumptions play an important
role.  For example, in the extreme case with only two groups and two periods, it is not clear whether the change over
time should be modeled as the same for the two groups in terms of levels of outcomes, or in terms of percentage changes in outcomes.  If the
initial period mean outcome is different across the two groups, the two different assumptions can give different answers in terms of both sign and
magnitude.  In general,
a treatment might affect both the mean and the variance of outcomes, and the impact of the treatment might vary across individuals.

For the case where the data includes repeated cross-sections of individuals (that is, the data include individual observations about
many units within each group in two different time periods, but the individuals can not be linked across time periods or may come
from a distinct sample), \citet{atheyimbens_cic} propose a non-linear
difference-in-difference model which they refer to as the changes-in-changes model that does not rely on functional form assumptions.

Modifying the notation from the last subsection, we now imagine that there are two groups, $g \in \{A,B\}$, where group A is the
control group and group B is the treatment group.  There are many individuals in each group with potential outcomes
denoted $Y_{gti}(w)$.  We observe $Y_{gti}(0)$ for a sample of units in both groups when $t=1$, and for group A when $t=2$;
we observe $Y_{gti}(1)$ for group B when $t=2.$ Denote the distribution of the observed outcomes in group $g$ at time $t$ by $F_{gt}(\cdot)$. We are interested
in the distribution of treatment effects for the treatment group in the second period, $Y_{B2i}(1)-Y_{B2i}(0)$. Note that
the distribution of $Y_{B2i}(1)$ is directly estimable, while the counterfactual distribution of $Y_{B2i}(0)$ is not, so the problem
boils down to learning the distribution of $Y_{B2i}(0)$, based on the distributions of 
$Y_{B1i}(0)$, $Y_{A2i}(0)$, and $Y_{A1i}(0)$.   Several assumptions
are required to accomplish this.  First is that the potential outcome in the absence of the treatment can be written as a monotone function of an unobservable $U_i$
and time: $Y_{gti}(0)=h(U_i,t)$.  Note that the function does not depend directly on $g$, so that differences across groups are 
attributed to differences in the distribution of $U_i$ across groups.  Second, the function $h$ is strictly increasing.  This is not a restrictive assumption for a single
time period, but it is restrictive when we require it to hold over time, in conjunction with a third assumption, namely that the 
distribution of $U_i$ is stable over time within each group.  The final assumption is that the support of $U_i$ for the treatment group
is contained in the support of $U_i$ for the control group.  Under these assumptions, the distribution of $Y_{B2i}(0)$ is identified,
with the formula for the distribution given as follows: 
\[ Pr(Y_{B2i}(0) \le y) = F_{B1}(F^{(-1)}_{A1}(F_{A2}(y))).  \]
\citet{atheyimbens_cic} show that an estimator based on the empirical distributions of the observed outcomes is efficient and discuss extensions to discrete outcome settings.

The nonlinear difference-in-difference model can be used for two distinct purposes.  First, the distribution is of direct interest
for policy, beyond the average treatment effect.  Further, a number of authors have used this approach as a robustness check, i.e.,
a supplementary analysis in the terminology of Section \ref{section:suppl}, for the results from a linear model.

\subsection{Estimating Average Treatment Effects under Unconfoundedness in Settings with Multivalued Treatments}
\label{subsection:multi}

Much of the earlier econometric literature on treatment effects focused on the case with binary treatments. For a textbook discussion, see  \cite{imbens2015causal}.
Here we discuss the results of the more recent multi-valued treatment effect literature. 
In the binary treatment case, many methods have been proposed for estimating the average treatment effect. Here we focus on 
two of these methods, subclassification with regression and and matching with regression, that have been found to be effective in the binary treatment case (\citet{imbens2015causal}). We discuss how these can be  extended to the multi-valued treatment setting without increasing the complexity of the estimators. In  particular, the dimension reducing properties of a generalized version of the propensity score can be maintained in the multi-valued treatment setting. 

\subsubsection{Set Up}
To set the stage, it is useful to start with the binary treatment case.
The standard set up postulates the existence of  two potential outcomes, $Y_i(0)$ and $Y_i(1)$. With the binary treatment denoted by $W_i\in\{0,1\}$, the realized and observed outcome is 
\[Y_i^\obs=Y_i(W_i)=\left\{\begin{array}{ll}
Y_i(0)\hskip1cm & {\rm if}\ W_i=0,\\
Y_i(1)\hskip1cm & {\rm if}\ W_i=1.\end{array}\right.
\]
In addition to the treatment indicator and the outcome we may observe a set of pretreatment variables denoted by $X_i$.
Following \citet{rosenbaum1983central} a large literature focused on estimation of the population average treatment effect
$\tau=\mme[Y_i(1)-Y_i(0)],$
under the unconfoundedness assumption
that
\[ W_i\ \indep\ \Bigl(Y_i(0),Y_i(1)\Bigr)\ \Bigl|\ X_i.\]
In combination with overlap, requiring that the propensity score
$e(x)={\rm pr}(W_i=1|X_i=x),$
is strictly between zero and one, the researcher can estimate the population average treatment effect by adjusting the differences in outcomes by treatment status for differences in the pretreatment variables:
\[ \tau=\mme\Bigl[
\mme[Y_i^\obs|X_i,W_i=1]-\mme[Y_i^\obs|X_i,W_i=0]
\Bigr].\]
In that case many  estimation strategies have been developed, relying on regression
\citet{hahn1998role}, matching \citet{abadie2006}, inverse propensity weighting \citet{hirr}, subclassification \citet{rosenbaum1983central}, as well as doubly robust methods \citet{robins1, robins2}. 
 \citet{rosenbaum1983central} established a key result that underlies a number of these estimation strategies: unconfoundedness implies that conditional on the propensity score, the assignment is independent of the potential outcomes:
\[ W_i\ \indep\ \Bigl(Y_i(0),Y_i(1)\Bigr)\ \Bigl|\ e(X_i).\]
In practice the most effective estimation methods appear to be those that combine some covariance adjustment through regression with a covariate balancing method such as subclassification, matching, or weighting based on the propensity score (\citet{imbens2015causal}).

Substantially less attention has been paid to the case where the treatment takes on multiple values. Exceptions include 
\cite{imbens2000, lechner2001, imai2004, cattaneo2010,  hirano2004} and \citet{yang2016}. 
 Let $\mathbb{W}=\{0,1,\ldots,T\}$ be the set of values for the treatment.
In the multivalued treatment case, one needs to be careful in defining estimands, and the role of the propensity score is subtly different. One natural set of estimands is
 the average treatment effect if all units were switched from treatment level $w_1$ to treatment level $w_2$:
\begin{equation}\label{est1} \tau_{w_1,w_2}=\mme[Y_i(w_2)-Y_i(w_1)].\end{equation}
To estimate  estimands corresponding to uniform policies such as (\ref{est1}), it is not sufficient to take all the units with treatment levels $w_1$ or $w_2$ and use methods for estimating treatment effects in a binary setting. The latter strategy would lead to an estimate of
$\tau'_{w_1,w_2}=\mme[Y_i(w_2)-Y_i(w_1)|W_i\in\{w_1,w_2\}]$, which differs in general from $\tau_{w_1,w_2}$ because of the conditioning. Focusing on unconditional average treatment effects like $\tau_{w_1,w_2}$ maintains transitivity:
$\tau_{w_1,w_2}+\tau_{w_2,w_3}=\tau_{w_1,w_3}$, which would not necessarily be the case for $\tau'_{w_1,w_2}$. 
There are other possible estimands, but we do not discuss alternatives here.

A  key first step is  to note that this estimand can be written as the difference in two marginal expectations:
$\tau_{w_1,w_2}=\mme[Y_{i}(w_2)]-\mme[Y_{i}(w_1)]$, and that therefore identification of marginal expectations such as $\mme[Y_{i}(w)]$ is sufficient for identification of average treatment effects.
 
Now suppose that a generalized version of unconfoundedness holds:
\[ W_i\ \indep\ \Bigl(Y_i(0),Y_i(1),\ldots, Y_i(T)\Bigr)\ \Bigl|\ X_i.\]
There is  no scalar function of the covariates  that maintains this conditional independence relation. 
In fact, with $T$ treatment levels one would need to condition on $T-1$ functions of the covariates to make this conditional independence hold.
However, unconfoundedness is in fact not required to enjoy the benefits of the dimension-reducing property of the propensity score.  \citet{imbens2000} introduces a  concept,  called weak unconfoundedness, which requires only that the indicator for receiving a particular level of the treatment and the potential outcome for that treatment level are conditionally independent:
\[ \mathbf{1}_{W_i=w}\ \indep\ Y_i(w)\ \Bigl|\ X_i, \ \ {\rm for\ all}\  w\in\{0,1,\ldots,T\}.\]
\cite{imbens2000}  shows that weak uncnfoundedness implies similar dimension reduction properties as are available in the binary treatment case.
He
further introduced the concept of the generalized propensity score:
\[ r(w,x)={\rm pr}(W_i=w|X_i=x).\]
Weak unconfoundedness implies that, for all $w$, it is sufficient for the removal of systematic biases to condition on the generalized propensity score for that particular treatment level:
\[ \mathbf{1}_{W_i=w}\ \indep\ Y_i(w)\ \Bigl|\ r(w,X_i).\]
This in turn can be used to develop matching or propensity score subclassification strategies as outlined in \citet{yang2016}.
This approach relies on the equality
$ \mme[Y_i(w)]
=\mme\Bigl[
\mme[Y_i^\obs|X_i,W_i=w]
\Bigr]$. As shown in  \citet{yang2016}, it follows from weak unconfoundedness that
\[ \mme[Y_i(w)]
=\mme\Bigl[
\mme[Y_i^\obs|r(w,X_i),W_i=w]
\Bigr].\]
To estimate $\mme[Y_i(w)]$, divide the sample into $J$ sublasses based on the value of $r(w,X_i)$, with $B_i\in\{1,\ldots,J\}$ denoting the subclass. We estimate
$\mu_j(w)=\mme[Y_i(w)|B_i=j]$ as the average of the outcomes for units with $W_i=w$ and $B_i=j$. Given those estimates, we estimate $\mu(w)=\mme[Y_i(w)]$ as a weighted average of the $\hat\mu_j(w)$, with weights equal to the fraction of units in subclass $j$.
The idea is not to find subsets of the covariate space where we can interpret the difference in average outcomes by all treatment levels as estimates of causal effects. Instead we find subsets where we can estimate the marginal average outcome for a particular treatment level as the conditional average for units with that treatment level, one treatment level at a time. This opens up the way for using matching and other propensity score methods developed for the case with binary treatments in settings with multivalued treatments, irrespective  of the number of treatment levels.

A separate literature has gone beyond the multi-valued treatment setting to look at dynamic treatment regimes. With few exceptions most of these studies appear in the biostatistical literature: see \citet{hernanrobins} for a general discussion.

\subsection{Causal Effects in Networks and Social Interactions}
\label{subsection:networks}

An important area that has seen much novel work in recent years is that on peer effects and causal effects in networks. Compared to the literature on estimating average causal effects unconfoundedness without interference, the literature has not focused on  a single setting;  rather, there are many problems and settings with interesting questions. Here, we will discuss some of the settings and some of the progress that has been made. However, this review will be brief, and incomplete, because this continues to be a very active area, with work ranging from econometrics (\citet{manski1993}) to economic theory (\citet{jackson}).

In general, the questions in this literature focus on causal effects in settings where units, often individuals, interact in a way that makes the no-interference or sutva (\citet{rosenbaum1983central, imbens2015causal})  assumptions that are routinely made in the treatment effect literature implausible.  Settings of interest  include those where the possible interference is simply a nuisance, and the interest continuous to be in causal effects of treatments assigned to a particular unit on the outcomes for that unit. There are also settings where the interest is in the magnitude of the interactions, or peer effects, that is, in the effects  of changing treatments  for one unit on the outcomes of other units. There are settings where the network (that is,  the set of links connecting the individuals) is fixed exogenously, and some where the network itself is the result of a possibly complex set of choices by individuals, possibly dynamic and possibly affected by treatments.
There are settings where the population can be partitioned into subpopulations with all units within a subpopulation connected, as, for example, in classroom settings (e.g., \citet{manski1993,carrell}), workers in a labor market (\citet{crepon})  or roommates in college (\citet{sacerdote}), or with general networks, where friends of friends are not necessarily friends themselves (\citet{christakis}). Sometimes it is more reasonable to think of many disconnected networks, where distributional approximations rely on the number of networks getting large, versus a single connected network such as Facebook. 
It maybe reasonable in some cases to think of the links as undirected (symmetric), and in others as directed. These links  can be binary, with links either present or not, or contain links of different strengths.
This large set of scenarios has led to the literature becoming somewhat fractured and unwieldy. We will only touch on a subset of these problems in this review.

\subsubsection{Models for Peer Effects}

Before considering estimation strategies, it is useful to begin by considering models of the outcomes in a setting with peer effects. Such models have been proposed in the literature.
A seminal paper in the econometric literature is Manski's linear-in-means model (\citet{manski1993, bramoulle, goldsmith}).
Manski's original paper focuses on the setting where the population is partioned into groups (e.g., classrooms), and peer effects are constant within the groups.  The basic model specification is
\[ Y_{i}=\beta_0+
\beta_\oy\cdot \oy_i+\beta_X'X_i+\beta_\ox'\ox_i+\beta_Z'Z_i+\varepsilon_i,\]
where $i$ indexes the individual.
Here $Y_{i}$ is the outcome for individual $i$, $\oy_i$ is the average outcome for individuals in the peer group for individual $i$, $X_i$ is a set of exogenous characteristics of individual $i$, $\ox_i$ is the average value of the characteristics in individual $i$'s peer group, and $Z_i$ are group characteristics that are constant for all individuals in the same peer group.
Manski considers three types of peer effects. Outcomes for individuals in the same group may be correlated because of a shared environment.  These effects are called correlated peer effects, and captured by the coefficient on $Z_i$.
Next are the exogenous peer effects, captured by the coefficient on the group average $\ox_i$ of the exogenous variables. The third type is the endogenous peer effect, captured by the coefficient on the group average outcomes $ \oy_i$.
Manski concludes that identification of these effects, even in the linear model setting, relies on very strong assumptions and is unrealistic in many settings. In subsequent empirical work, researchers have often ruled out some of these effects in order to identify others.

\citet{graham_ectrica} focuses on a setting very similar to that of Manski's linear-in-means model. He considers restrictions on the covariance matrix within peer groups implied by the model assuming homoskedasticity at the individual level.
\citet{bramoulle} allows for a more general network configuration than Manski, and  investigate the benefits of such configurations for identification in the Manski-style linear-in-means model. 
\citet{hudgens} start closer to the Rubin Causal Model or potential outcome setup. Like Manski they focus on a setting with a partitioned network. Following the treatment effect literature they focus primarily on the case with a binary treatment. Let $W_i$ denote the treatment for individual $i$, and
let $\bw_i$ denote the vector of treatments for the peer group for individual $i$. The starting point in the 
\citet{hudgens}  set up  is the potential outcome $Y_i(\bww)$,  with restrictions placed on the dependence of the potential outcomes on the full treatment vector $\bww$.
\citet{aronowsamii} allow for general networks and peer effects, investigating the identifying power from randomization.

\subsubsection{Models for Network Formation}

Another part of the literature has focused on developing models for network formation. Such models are of interest in their own right, but they are also important for deriving asymptotic approximations based on large samples. Such approximations require the researcher to specify in what way the expanding sample would be similar to or different from the current sample. For example, it would require the researcher to be specific in the way the additional units would be linked to current units or other new units.

There is a wide range of models considered, with some models relying more heavily on optimizing behavior of individuals, and others using more statistical models. See \citet{goldsmith,christakisimbens,mele,jackson,jacksonwolinsky} for such network models in economics, and \citet{hollandleinhardt} for statistical models.  \citet{arun2015network} develops a model for network formation and develops a corresponding central limit theorem in the presence of correlation induced by network links.  \citet{arun2015econometrics}
surveys the econometrics of network formation.

\subsubsection{Exact Tests for Interactions}

One challenge in testing hypotheses about peer effects using methods based on standard asymptotic theory is that when individuals interact (e.g., in a network), it is not clear how interactions among individuals would change as the network grows. Such a theory would require a model for network formation, as discussed in the last subsection.
This motivates an approach that allows us to test hypotheses without invoking large sample properties of test statistics (such as asymptotic normality). Instead, the distributions of the test statistics are based on the random assignments of the treatment, that is, the properties of the tests are based on randomization inference.  In randomization inference, we approximate the distribution of the test statistic under the null 
hypothesis by re-calculating the test statistic under a large number of alternative (hypothetical) treatment assignment vectors, where the alternative treatment assignment vectors are drawn from the randomization
distribution.  For example, if units were independently assigned to treatment status with probability $p$, we re-draw hypothetical assignment vectors with each unit assigned to treatment
with probability $p$.  Of course, re-calculating the test statistic requires knowing the values of units' outcomes.  The randomization inference approach is easily applied if the null
hypothesis of interest is ``sharp'': that is, the null hypothesis specifies what outcomes would be under all possible treatment assignment vectors.  If the null hypothesis is that the 
treatment has no effect on any units, this null is sharp: we can infer what outcomes would have been under alternative treatment assignment vectors, in in particular, outcomes would
be the same as the realized outcomes under the realized treatment vector.
  
More generally, however, randomization inference for tests for peer effects is more complicated than in settings without peer effects because the null hypotheses are often {\it not} sharp. 
\citet{aronow2012,atheyeckles}
 develop methods for calculating exact p-values for general null hypotheses on causal effects in a single connected network, allowing for peer effects. The basic case \citet{aronow2012,atheyeckles}
 consider is that where the null hypothesis rules out peer effects but allows for direct (own) effects of a binary treatment assigned randomly at the individual level.  Given that direct effects
 are not specified under the null, individual outcomes are not known under alternative treatment assignment vectors, and so the null is not sharp.  To address this problem,
\citet{atheyeckles} introduce the notion of an artificial experiment that differs from the actual experiment. In the artificial experiment, some units have their treatment assignments held
fixed, and we randomize over the remaining units.  Thus, the randomization distribution is replaced by a conditional randomization distribution, where treatment assignments of some units are re-randomized
conditional on the assignment of other units.  By focusing on the conditional assignment given a subset of the overall space of assignments, and by focusing on outcomes for a subset of the units in the original experiment, they create an artificial experiment where the original null hypothesis that was not sharp in the original experiment is now sharp.
To be specific, the artificial experiments starts by designating an arbitrary set of units to be focal. The test statistics considered depend only on outcomes for these focal units.
Given the focal units, the set of assignments that, under the null hypothesis of interest, does not change the outcomes for the focal units is derived. The exact distribution of the test statistic can then be inferred for such test statistics under that conditional randomization distribution under the null hypothesis considered.   

\citet{atheyeckles}
 extend this idea to a large class of null hypotheses. This class includes hypotheses restricting higher order peer effects (peer effects from friends-of-friends) while allowing for the presence of peer effects from friends. It also includes hypotheses about the validity of sparsification of a dense network, where the question concerns peer effects of friends according to the pre-sparsified network while allowing for peer effects of the sparsified network. Finally, the class also includes null hypotheses concerning the exchangeability of peers. In many models peer effects are restricted so that all peers have equal effects on an individual's outcome. It may be more realistic to allow effects of highly connected individuals, or closer friends, to be be different from those of less connected or more distant friends. Such hypotheses can be tested in this framework.

\subsection{Randomization Inference and Causal Regressions}
\label{subsection:regression}

In recent empirical work,
data from randomized experiments are often analyzed using conventional regression methods. Some researchers have raised concerns with  the regression approach in small samples (\citet{freedman1, freedman2, young, atheyimbens2016, imbens2015causal}), but generally such analyses are justified at least in large samples, even in settings with many covariates 
(\citet{binyu, tibs}). There is an alternative approach to estimation and inference, however, that does not rely on large sample approximations, using approximations for the distribution of estimators induced by randomization. 
Such methods, which go back to
\citet{fisher1925, fisher1935, neyman1923, neyman1935},
 clarify how the act of randomization allows for the testing for the presence of treatment effects and the unbiased estimation of average treatment effects. Traditionally these methods have not been used much in economics.
However, recently there has been some renewed interest in such methods. See for example \citet{imbensrosenbaum, young, atheyimbens2016}).  In completely randomized experiments these methods are often straightforward, although even there analyses involving covariates can be more complicated.

However, the value of the randomization perspective extends well beyond the analysis of actual experiments. It can shed light on the interpretation of observational studies and the complications arising from finite population inference and clustering. Here we discuss some of these issues and more generally provide an explicitly causal perspective on linear regression.
Most textbook discussions of regression specify the regression function in terms of a dependent  variable,  a number of explanatory variables, and an unobserved component, the latter often referred to as the error term:
\[ Y_i=\beta_0+\sum_{k=1}^K \beta_k\cdot X_{ik}+\varepsilon_i.\]
Often the assumption is made that in the population the units are randomly sampled from, the unobserved component $\varepsilon_i$ is independent of, or uncorrelated with, the regressors $X_{ik}$. The regression coefficients are then estimated by least squares, with the uncertainty in the estimates interpreted as sampling uncertainty induced by random sampling from the large population.

This approach works well in many cases. In analyses using data from the public use surveys such as the Current Population Survey or the Panel Study of Income Dynamics it is natural to view the sample at hand as a random sample from a large population. In other cases this perspective is not so natural, with the sample not drawn from a well-defined population. This includes convenience samples, as well as settings where we observe all units in the population. In those cases it is helpful to take an explictly causal perspective. This perspective also clarifies how the assumptions underlying identification of causal effects relate to the assumptions often made in least squares approaches to estimation.

Let us separate the covariates $X_i$ into a subset of causal variables $W_i$ and the remainder, viewed as fixed characteristics of the units. For example, in a wage regression the causal variable may be years of education and the characteristics may include sex, age, and parental background. Using the potential outcomes perspective we can interpret $Y_i(w)$ as the outcome corresponding to a level of the treatment $w$ for unit or individual $i$.
Now suppose that for all units $i$ the function $Y_i(\cdot)$ is linear in in its argument, with a common slope coefficient, but a variable intercept, $Y_i(w)=Y_i(0)+\beta_W\cdot w$. Now write $Y_i(0)$, the outcome for unit $i$ given treatment level $0$ as
\[ Y_i(0)=\beta_0+\beta_Z'Z_i+\varepsilon_i,\]
where $\beta_0$ and $\beta_Z$ are the population best linear predictor coefficients. This representation of $Y_i(0)$ is purely definitional and does not require assumptions on the population.
Then we can write the model as
\[ Y_i(w)=\beta_0+\beta_W\cdot w+\beta_Z'Z_i+\varepsilon_i,\]
and the realized outcome as
\[ Y_i=\beta_0+\beta_W\cdot W_i+\beta_Z'Z_i+\varepsilon_i.\]
Now we can investigate the properties of the least squares estimator 
$\hat \beta_W$ for $\beta_W$, where the distribution of $\hat\beta_W $ is generated by the assignment mechanism for the $W_i$. In the simple case where there are no characteristics $Z_i$ and the cause $W_i$ is a binary indicator, the assumption that the cause is completely randomly assigned leads to the conventional Eicker-Huber-White standard errors (\citet{eicker, huber, white1980robust}). Thus, in that case viewing the randomness as arising from the assignment of the causes rather than as sampling uncertainty provides a coherent way of interpreting the uncertainty.

This extends very easily to the case where $W_i$ is binary and completely randomly assigned but there are other regressors included in the regression function. As  \citet{lin} and \citet{imbens2015causal} show there is no need for assumptions about the relation of those regressors to the outcome, as long as the cause $W_i$ is randomly assigned.
\citet{abadieathey} extend this to the case where the cause is multivalued, possibly continuous, and the characteristics $Z_i$ are allowed to be generally correlated with the cause $W_i$. 
\citet{aronowsamii} discuss the interpretation of the regression estimates in a causal framework.
\citet{abadieathey2} discuss extensions to settings with clustering where the need for clustering adjustments in standard errors arises from the clustered assignment of the treatment rather than through clustered sampling.

\subsection{External Validity}
\label{subsection:external}

One concern that has been raised in many studies of causal effects is that of external validity. Even if a causal study is done carefully, either in analysis or by design, so that the internal validity of such a study is high, there is often little guarantee that the causal effects are valid for populations or settings other than those studied.  This concern has been raised particularly forcefully in experimental studies where the internal validity is guaranteed by design. See for example the discussion in \citet{deaton, imbens2010} and \citet{manski2013public}.
Traditionally, there has been much emphasis on internal validity in studies of causal effects, with some arguing for the primacy of internal validity.  Some have argued that without internal validity, little can be learned from a study (\citet{shadishcookcampbell}, \citet{imbensmanski}).  Recently, however,  \citet{deaton, manski2013public, banerjee} have argued that external validity should receive more emphasis. 

Some recent work has taken concerns with external validity more seriously, proposing a variety of approaches that directly allow researchers to assess the external validity of estimators for causal effects. A leading example concerns settings with instrumental variables with heterogenous treatment effects (e.g., \citet{angrist2004, angristfernandez, donglewbel, angristrokkanen, bertanhaimbens, kowalski, mogstad}).  In the modern literature
 with heterogenous treatment effects  
the instrumental variables estimator is interpreted as an estimator of the local average treatment effect,  the average effect of the treatment for the compliers, that is, individuals whose treatment status is affected by the instrument.
In this setting, the focus has been on whether the instrumental variables estimates are relevant for the entire sample, that is, have external validity, or only have local validity for the complier subpopulation.

 In that context, \citet{angrist2004} suggests testing whether the difference in average outcomes for always-takers and never-takers is equal to the average effect for compliers. In this 
 context, a Hausman test \citep{hausman} for equality of the ordinary least squares estimate and an instrumental variables estimate can be interpreted as testing whether the average treatment effect is equal to the local average treatment effect; of course, the ordinary least squares estimate only has that interpretation if unconfoundedness holds. \citet{bertanhaimbens} suggest testing a combination of two equalities, first that the average outcome for untreated compliers  is equal to the average outcome for never-takers, and second, that the average outcome for treated compliers is equal to the average outcome for always-takers. This turns out to be equivalent to testing both the null hypothesis suggested by  \citet{angrist2004} and the Hausman null.
\citet{angristfernandez} consider extrapolating local average treatment effects by exploiting the presence of
other exogenous covariates. The key assumption in the  
\citet{angristfernandez} approach,  ``conditional effect ignorability,'' is that conditional
on these additional covariates the average effect for compliers is identical to the average effect for never-takers and always-takers.

In the context of regression discontinuity designs, and especially in the fuzzy regression discontinuity setting, the concerns about external validity are especially salient. In that setting the estimates are in principle valid only for individuals with values of the forcing variable equal to, or close to, the threshold at which the probability of receipt of the treatment changes discontinuously. There have been a number of approaches to assess the plausibility of generalizing those local estimates to other parts of the population.
The focus and the applicability of the various methods to assess external validity varies. Some of them apply to both sharp and fuzzy regression discontinuity designs, and some apply only to fuzzy designs.
Some require the presence of additional exogenous covariates, and others rely only on the presence of the forcing variable. \citet{donglewbel} observe that in general, in regression discontinuity designs with a continuous forcing variable,
 one can estimate the magnitude of the discontinuity as well as the magnitude of the change in the first 
 derivative of the regression function, or even higher order derivatives. Under assumptions about the smoothness of the two conditional mean functions, knowing
 the higher order derivatives allows one to extrapolate  away from values of the forcing variable close to the threshold.
This method apply both in the sharp and in the fuzzy regression discontinuity design. It does not require the presence of additional covariates.  In another approach,
\citet{angristrokkanen} do require the presence of additional exogenous covariates. They suggest testing whether whether conditional
on these covariates, the correlation between the forcing variable and the outcome vanishes. This would imply that the assignment can be thought of as unconfounded conditional on the additional covariates. Thus it would allow for extrapolation away
from the threshold. Like the Dong-Lewbel approach, the  Angrist-Rokkanen methods apply both in the case of sharp and fuzzy regression discontinuity designs.
Finally, \citet{bertanhaimbens} propose an approach requiring a fuzzy regression discontinuity design.  They suggest testing for continuity of the conditional expectation of the outcome conditional on the treatment and the forcing variable, at the threshold, adjusted for differences in the covariates.

\subsection{Leveraging Experiments}
\label{subsection:leveraging}

Randomized experiments are the most credible design to learn about causal effects. However, in practice there are often reasons that researchers cannot conduct randomized experiments to answer the causal questions of interest. They may be expensive, or they may take too long to give the researcher the answers that are needed now to make decisions, or there may be ethical objections to experimentation. As a result, we often rely on a combination of experimental results and observational studies to make inferences and decisions about a wide range of questions. In those cases we wish to exploit the benefits of the experimental results, in particular the high degree of internal validity, in combination with the external validity and precision from large scale representative observational studies. At an abstract level, the observational data are used to estimate rich models that allow one to answer many questions, but the model is forced to accommodate the answers from the experimental data for the limited set of questions the latter can address. Doing so will improve the answers from the observational data without compromising their ability to answer more questions.

Here we discuss two specific settings where experimental studies can be leveraged in combination with observational studies to provide richer answers than either of the designs could provide on their own.
In both cases, the interest is in the average causal effect of a binary treatment on a primary outcome. However, in the experiment the primary outcome was not observed and so one cannot directly estimate the average effect of interest. Instead an intermediate outcome was observed. In a second study, both the intermediate outcome and the primary outcome were observed. In both studies there may be additional pretreatment variables observed and possibly the treatment indicator.

These two examples do not exhaust the set of possible settings where researchers can leverage experimental data more effectively, and this is likely to be an area where more research is fruitful.

\subsubsection{Surrogate Variables}
In the first setting, studied in  \citet{atheychetty}, in the second  sample the treatment indicator  is not observed. In this case researchers may wish to use the intermediate variable, denoted $S_i$, as a surrogate. Following \citet{prentice, begg2000, frangakisrubin}, the key condition for an intermediate variable to be a surrogate is that in the experimental sample, conditional the surrogate and observed covariates, the (primary) outcomes and the treatment are independent: $Y_i \indep W_i | (S_i, X_i)$. There is a long history of attempts to use intermediate health measures in medical trials as surrogates (\citet{prentice}). The results are mixed, with the condition often not satisfied in settings where it could be tested. However, many of these studies use low-dimensional surrogates. In modern settings there is often a large number of intermediate variables recorded in administrative data bases that lie on or close to the causal path between the treatment and the primary outcome. In such cases it may be more plausible that the full set of surrogate variables satisfies at least approximately the surrogacy condition.

For example, suppose an internet company is considering a change to the user experience on the company's website. They are interested in the effect of that change on the user's engagement with the website over a year long period. They carry out a randomized experiment over a month, where they measure details about the user's engagement, including the number of visits, webpages visited, and the length of time spent on the various webpages. In addition, they may have historical records on user characteristics including past engagement, for a large number of users. The combination of the pretreatment variables and the surrogates may be sufficiently rich so that conditional on the combination the primary outcome is independent of the treatment.

Given surrogacy, and given comparability of the observational and experimental sample (which requires that the conditional distribution of the primary outcome given surrogates and pretreatment variables is the same in the experimental and observational sample), \citet{atheychetty} develop two methods for estimating the average effect. The first corresponds to estimating the relation between the outcome and the surrogates in the observational data and using that to impute the missing outcomes in the experimental sample. The second corresponds to estimating the relation between the treatment and the surrogates in the experimental sample and use that to impute the treatment indicator in the observational sample. They also derive the biases from violations of the surrogacy assumption. 

\subsubsection{Experiments and Observational Studies}

In the second setting, studied in \citet{atheychettyimbens},
the researcher again has data from a randomized experiment containing information on the treatment and the intermediate variables, as well as pretreatment variables. In the observational study the researcher now observes the same variables plus the primary outcome. If in the observational study unconfoundedness (selection-on-observables) were to hold, the researcher would not need the experimental sample, and could simply estimate the average effect of the treatment on the primary outcome by adjusting for differences between treated and control units in pretreatment variables. However, one can compare the estimates of the average effect on the intermediate outcomes based on the observational sample, after adjusting for pretreatment variables, with those from the experimental sample. The latter are known to be consistent, and so if one finds substantial and statistically significant differences, unconfoundedness need not hold. For that case \citet{atheychettyimbens} develop methods for adjusting for selection on unobservables exploiting the observations on the intermediate variables.

\subsubsection{Multiple Experiments}

An issue that has not received as much attention, but provides fertile ground for future work concerns the use of multiple experiments. Consider a setting where a number of experiments were conducted. The experiments may vary in terms of the population that the sample is drawn from, or in the exact nature of the treatments included. The researcher may be interested in combining these experiments to obtain more efficient estimates, predicting the effect of a treatment in another population, or estimating the effect of a treatment with different characteristics. Such inferences are not validated by the design of the experiments, but the experiments are important in making such inferences more credible. These issues are related to external validity concerns, but include more general efforts to decompose experimentally estimated effects into components that can inform decisions on related treatments.
In the treatment effect literature aspects of these problems have been studied in
\citet{hotzimbensmortimer,imbens2010,alcott}. They have also received some attention in the literature on structural modeling, where the experimental data are used to anchor aspects of the structural model, e.g., \citet{toddwolpin}.

\section{Supplementary Analyses}
\label{section:suppl}

One common feature of much of the empirical work in the causal literature is the use of what we call here {\it supplementary analyses}. 
We want to contrast supplementary analyses with primary analyses whose focus is on point estimates of the primary estimands and standard errors thereof. Instead, the point of the
supplementary analyses  is to shed light on the credibility of the primary analyses.
They are intended to probe the identification strategy underlying the primary analyses. The results of these supplementary analyses is not to end up with a better estimate of the effect of primary interest. The goal is also not to directly select among competing statistical models. Rather, the results of the supplementary analyses  either enhance the credibility of the primary analyses or cast doubts on them, without necessarily suggesting alternatives to these primary analyses (although sometimes they may).  The supplementary analyses are often based on  careful and creative examinations of the identification strategy. Although at first glance, this creativity may appear
application-specific, in this section we try to highlight some common themes.

 In general, the assumptions behind the identification strategy often have implications for the data beyond those exploited in the primary analyses, and these implications are the focus of the supplementary analyses. The supplementary analyses can take on a variety of forms, and we are currently not aware of a comprehensive survey to date.
Here we discuss some examples from the empirical and theoretical literatures and draw some general conclusions in the hope of providing some guidance for future work. This is a very active literature, both in theoretical and empirical studies, and it is likely that the development of these methods will continue rapidly.

The assumptions underlying identification strategies can typically be stated without reference to functional form
assumptions or estimation strategies.  For example, unconfoundedness is a conditional independence assumption. 
There are variety of estimation strategies that exploit the unconfoundedness assumption.  Supplementary analyses may
attempt to establish the credibility of the underlying independence assumption; or, they may jointly establish the credibility
of the underlying assumption and the specific estimation approach used for the primary analysis. 

In Section \ref{subsection:placebo} we discuss one of the most common forms of supplementary analyses, placebo analysis, where pseudo causal effects are estimated that are known to be equal to zero. In Section \ref{subsection:robustness}
we discuss sensitivity and robustness analyses that assess how much estimates of the primary estimands can change if we weaken the critical assumptions underlying the primary analyses.
In Section \ref{subsection:identification} we discuss some recent work on understanding the identification of key model estimates by linking model parameters to summary statistics of the data.
In Section \ref{subsection:suppl_rdd} we disuss a particular supplementary analysis that is specific to regression discontinuity analyses. In this case the focus is on the continuity of the density of an exogenous variable, with a discontinuity as the threshold for the regression discontinuity analyses evidence of manipulation of the forcing variable.

\subsection{Placebo Analyses}
\label{subsection:placebo}

The most widely used of the supplementary analyses is what is often referred to as a placebo analysis. In this case the researcher replicates the primary analysis with the outcome replaced by a pseudo outcome that is known not to be affected by the treatment. Thus, the true value of the estimand for this pseudo outcome is zero, and the goal of the supplementary analysis is to assess whether the adjustment methods employed in the primary analysis when applied to the pseudo outcome lead to estimates that are close to zero, taking into account the statistical uncertainty.
Here we discuss some settings where such analyses, in different forms, have been applied, and provide some general guidance. Although these analyses often take the form of estimating an average treatment effect and testing whether that is equal to zero, underlying the approach is often conditional independence relation.  In this review we highlight the fact that there is typically more to be tested than simply a single average treatment effect.

\subsubsection{Lagged Outcomes}

One type of placebo test relies on treating lagged outcomes as pseudo outcomes. Consider, for example, the lottery data set assembled by \citet{imbensrubinsacerdote}, which studies participants in the Massachusetts state lottery. The treatment of interest is an indicator for winning a big prize in the lottery (with these prizes paid out over a twenty year period), with the control group consisting of individuals who won one small, one-time prizes. The estimates of the average treatment effect rely on an unconfoundedness assumption, namely that the lottery prize is as good as randomly assigned after taking out associations with some pre-lottery variables:
\begin{equation}\label{unconf}
 W_i\ \indep\ \Bigl(Y_i(0),Y_i(1)\Bigr) \Bigl|\ X_i.\end{equation}
The pre-treatment variables include six years of lagged earnings as well as six individual characteristics (including education measures and gender).
 Unconfoundedness is plausible here because  which ticket wins the lottery is random, but because of a 50\% response rate, as well as differences in the rate at which individuals buy lottery tickets, there is no guarantee that this assumption holds. To assess the assumption it is useful to estimate the same regression function with pre-lottery earnings as the outcome, and the indicator for winning on the right hand side with the same set of additional exogenous covariates.
Formally,
we partition the vector of covariates $X_i$ into two parts, a (scalar) pseudo outcome, denoted by $\xp_i$, and the remainder, denoted by  $\xr_i$, so that
$X_i=(\xp_i,\xr_i)$. We can then test the conditional independence relation
\begin{equation}\label{indep}
W_i\ \indep\  \xp_i\ \Bigl|\ \xr_i.\end{equation}
Why is testing this conditional independence relation relevant for assessing unconfoundedness in (\ref{unconf})? There are two conceptual steps. One is that the pseudo outcome $\xp_i$ is viewed as a proxy for one or both of the potential outcomes. Second, it relies on the notion that if unconfoundedness holds given the full set of pretreatment variables $X_i$, it is plausible that it also holds given the subset $\xr_i$. In the lottery application, taking $\xp_i$ to be earnings in the year prior to winning or not, both steps appear plausible.
Results for this analysis are in Table \ref{tabel_lottery}. 
Using the actual outcome we estimate that winning the lottery (with on average a \$20,000 yearly prize), reduces average post-lottery earnings by \$5,740, with a standard error of \$1,400. Using the pseudo outcome we obtain an estimate of minus  \$530, with a standard error of \$780.

\begin{table}
 \caption{\sc  Lagged as Pseudo-Outcomes in the Lottery Data}
 \vskip1cm
\begin{center}
\begin{tabular}{lccccc}
  Outcome  & est & (s.e.) \\
\hline\\
Pseudo Outcome: $Y_{-1,i}$  &  -0.53  &  (0.78)\\ \\
Actual Outcome: $Y^\obs_i$  & {-5.74} & {(1.40)}\\
\end{tabular}      
\end{center}      \label{tabel_lottery}
\end{table}

In Table \ref{tabel_lottery2}, we take this one step further by testing the conditional independence relation in (\ref{indep}) more fully. We do this by testing the null of no average difference for two functions of the pseudo-outcome, namely the actual level and an indicator for the pseudo-outcome being positive. Moreover we test this separately for individuals with positive earnings two years prior to the lottery and individuals with zero earnings two years prior to the lottery. Combining these four tests in a chi-squared statistic leads to a p-value of 0.135. 
Overall these analyses are supportive of unconfoundedness holding in this study.

\begin{table}
 \caption{\sc  Testing Conditional Independence of Lagged Outcomes and the Treatment in the Lottery Data}
 \vskip1cm
\begin{center}
\begin{tabular}{lccccccc}
\hline\hline
Pseudo  & Subpopulation & est & (s.e.) \\
Outcome \\
\hline
& \\
$\mathbf{1}_{\{Y_{-1,i}=0\}}$ & $Y_{-2,i}=0 $  & -0.07 & (0.78) \\ 
$\mathbf{1}_{\{Y_{-1,i}=0\}}$ & $Y_{-2,i}>0$  & 0.02 & (0.02) \\ 
$Y_{-1,i}$ & $Y_{-2,i}=0 $ &  -0.31 & (0.30)  \\ 
$Y_{-1,i}$ & $Y_{-2,i}>0 $  &0.05 & (0.06)\\ 
\\
 & &statistic & p-value \\
Combined Statistic & \\
(chi-squared, dof 4)&  & 2.20 & 0.135\\
\hline 
\end{tabular}      
\end{center}    \label{tabel_lottery2}
\end{table}

Using the same approach with the \citet{lalonde} data that are widely used in the evaluation literature (e.g., \citet{heckmanhotz, dehejiawahba, imbens2015}), the results are quite different. Here we use 1975 earnings as the pseudo-outcome, leaving us with only a single pretreatment year of earnings to adjust for the substantial difference between the trainees and comparison group from the CPS. Now, as reported in Table \ref{tabel_lalonde}, the adjusted differences between trainees and CPS controls remain substantial, casting doubt on the unconfoundedness assumption. Again we first test whether the simple average difference in adjusted 1975 earnings is zero. Then we test whether both the level of 1975 earnings and the indicator for positive 1975 earnings are different in the two groups, separately for individuals with zero and positive 1974 earnings.  The null is rejected, casting doubt on the unconfoundedness assumption (together with the
approach for controlling for covariates, in this case subclassification).

\begin{table}[ht]
 \caption{\sc  Lagged Earnings as a Pseudo-Outcome in the Lalonde Data}
 \vskip1cm
\begin{center}
\begin{tabular}{lccc}
\\
\hline\hline
& & & p-value
\\ \\ 
earnings\ 1975: &  -0.90  & (0.33) & 0.006\\
 \\
chi-squared test  & 53.8 & (dof=4) & $<0.001$ \\
\\
\hline
\end{tabular}
\end{center}\label{tabel_lalonde}
\end{table}

\subsubsection{Covariates in Regression Discontinuity Analyses}

As a second example, consider a regression discontinuity design. Covariates typically play only a minor role in the primary analyses there, although they can improve precision (\citet{imbenslemieux, calonico, calonico2}). The reason is that in most applications of regression discontinuity designs, the covariates are uncorrelated with the treatment conditional on the forcing variable being close to the threshold. As a result, they are not required for eliminating bias.
However, these exogenous covariates can play an important role  in assessing the plausibility of the design. According to the identification strategy, they should be uncorrelated with the treatment when the forcing variable is close to the threshold. However, there is nothing in the data that guarantees that this holds. We can therefore test this conditional independence, for example by using a covariate as the pseudo outcome in a regression discontinuity analysis. If we were to find that the conditional expectation of one of the covariates is discontinuous at the threshold, it would cast doubt on the identification strategy. Note that formally, we do not need this conditional independence to hold, and if it were to fail one might be tempted to simply adjust for it in a regression analysis. However, the presence of such a discontinuity may be difficult to explain in a regression discontinuity design, and adjusted estimates would therefore not have much credibility.  The 
discontinuity might be interpreted as evidence for an unobserved confounder whose distribution changes at the boundary, one which might also be correlated with the outcome of interest.

Let us illustrate this with the Lee election data (\citet{lee}). 
\citet{lee} is interested in estimating the effect of incumbency on electoral outcomes.
The treatment is a Democrat winning a congressional election, and the forcing variable is the Democratic vote share minus the Republication vote share in the current election.  We look at an indicator for winning the next election as the outcome.  As a pretreatment variable, we consider an indicator for winning the previous election to the one that defines the forcing variable. Table \ref{tabel_lee} presents the results, based on the Imbens-Kalyanaraman bandwidth, where we use local linear regression (weighted with a triangular kernel to account for boundary issues).
\begin{table}[ht]
 \caption{\sc  Winning a Previous Election as a Pseudo-Outcome in Election Data}
 \vskip1cm
\begin{center}
\begin{tabular}{lccc}
\\
\hline\hline
\\ \\ 
Democrat Winning Next Election&  0.43  & (0.03) & 0.26 \\
 \\
Democrat Winning Previous Election & 0.03 & (0.03)  & 0.19\\
\\
\hline
\end{tabular}
\end{center}\label{tabel_lee}
\end{table}
The estimates for the actual outcome (winning the next election) are substantially larger than those for the pseudo outcome (winning the previous election), where we cannot reject the
null hypothesis that the effect on the pseudo outcome is zero.


\subsubsection{Multiple Control Groups}

Another example of the use of placebo regressions is \citet{rosenbaum_multiple} (see also \citet{heckmanhotz, imbens2015causal}). \citet{rosenbaum_multiple} is interested in the causal effect of a binary treatment and focuses on a setting with multiple comparison groups. There is no strong reason to believe that one of the comparison groups is superior to another. \citet{rosenbaum_multiple} proposes testing equality of the average outcomes in the two comparison groups after adjusting for pretreatment variables. If one finds that there are substantial differences left after such adjustments, it shows that at least one of the comparison groups is not valid, which makes the use of either of them less credible.  In applications to evaluations of labor market programs one might implement such methods by comparing individuals who are eligible but choose not to participate, to individuals who are not eligible. The biases from evaluations based on the first control group might correspond to differences in motivation, whereas evaluations based on the second control group could be biased because of direct associations between eligibility criteria and outcomes.

Note that one can also exploit the presence of multiple control groups by comparing estimates of the actual treatment effect based on one comparison group to that based on a second comparison group. Although this approach seems appealing at first glance, it is in fact less effective than direct comparisons of the two comparison groups because comparing treatment effect estimates involves the data for the treatment group, whose outcomes are not relevant for the hypothesis at hand.

\subsection{Robustness and Sensitivity}
\label{subsection:robustness}

Another form of supplementary analyses focuses on sensitivity and robustness measures.
The classical frequentist statistical paradigm suggests that a researcher specifies a single statistical model. The researcher then estimates this model on the data, and reports estimates and standard errors. The standard errors and the corresponding confidence intervals are valid given under the assumption that the model is correctly specified, and estimated only once. This is of course far from common practice, as pointed out, for example, in \citet{leamer_book, leamer_aer}. 
In practice researcher consider many specifications and perform various specification tests before settling on a preferred model. Not all the intermediate estimation results and tests are reported.

A common practice in modern empirical work is to present in the final paper estimates of the preferred specification  of the 
model, in combination with  assessments of the robustness of the findings from this preferred specification. These alternative specifications are not intended to be interpreted as statistical tests of the validity of the preferred model, rather they are intended to convey that the substantive results of the preferred specification are not sensitive to some of the choices in that specification.
These alternative specifications may involve different functional forms of the regression function, or different ways of controlling for differences in subpopulations.
Recently there has been some work trying to make these efforts at assessing robustness more systematic.

\citet{atheyimbens_robust} propose an approach to this problem.  We can illustrate the approach in the context of regression analyses, although it can also be applied to more complex nonlinear or structural models. In the regression context, suppose that the object of interest is a particular regression coefficient that has an interpretation as a causal effect. For example, in the preferred specification
\[ \mme[Y_i|X_i,Z_i]=\beta_0+\beta_W\cdot W_i+\beta_Z'Z_i,\]
the interest may be in $\beta_W$, the coefficient on $W_i$. They then suggest considering a set of different specifications based on splitting the sample into two subsamples, with $X_i\in\{0,1\}$ denoting the subsample, and in each case estimating
\[ \mme[Y_i|W_i,Z_i,Z_i=z]=\beta_{0x}+\beta_{Wx}\cdot W_i+\beta_{Zx}'Z_i.\]
The original causal effect is then estimated as $\tilde\beta_W=\overline{X}\cdot\hat\beta_{W1}+
(1-\overline{X})\cdot\hat\beta_{W0}$.
If the original model is correct, the augmented model still leads to a consistent estimator for the estimand.
 \citet{atheyimbens_robust} suggest splitting the original sample once for each of the elements of the original covariate vector $Z_i$, and splitting at a threshold that optimizes fit by minimizing the sum of squared residuals. Note that the focus is not on finding an alternative specification that may provide a better fit; rather, it is on assessing whether the estimate in the original specification is robust to a range of alternative specifications. They suggest reporting the standard deviation of the $\tilde\beta_W$ over the set of sample splits, rather than the full set of estimates for all sample splits.  This approach has some weaknesses, however.  For example, adding irrelevant covariates
 to the procedure might decrease the standard deviation of estimates.  If there are many covariates, some form of dimensionality 
 reduction may be appropriate prior to estimating the robustness measure.  Refinements and improvements on this approach is
 an interesting direction for future work.

Another place where it is natural to assess robustness is in estimation of average treatment effects $\mme[Y_i(1)-Y_i(0)]$ under unconfoundedness or selection on observables,
\[ W_i\ \indep\ \Bigl(Y_i(0),Y_i(1)\Bigr)\ \Bigl| X_i.\]
The theoretical literature has developed many estimators in the setting with unconfoundedness. Some rely on estimating
the conditional mean, $\mme[Y_i|X_i,W_i]$, some rely on estimating the propensity score $\mme[W_i|X_i]$, while others rely on matching on the covariates or the propensity score. See \citet{imbenswooldridge} for a review of this literature.
We believe that researchers should not rely on a single method, but report estimates
 estimation based on a variety of  methods to assess robustness.

\citet{arkhangelsky}  studies sensitivity of the estimates of the parameters of interest to misspecification of the model governing the nuisance parameters.
Another way to assess robustness is to use the partial indentification or bounds literature originating with \citet{manski_bounds}. See \citet{tamer} for a recent review. In combination with reporting estimates based on the preferred specification that may lead to point identification, it may be useful to combine that with reporting ranges based substantially weaker assumptions.
Coming at the same problem as the bounds approach, but from the opposite direction, 
\citet{rosenbaumrubin_sensitivity,rosenbaum_book} suggest sensitivity analyses. Here the idea is to start with a restrictive specification, and to assess the changes in the estimates that result from small to modest relaxations of the key identifying assumptions such as unconfoundedness. In the context  
\citet{rosenbaumrubin_sensitivity}
 consider, that of estimating average treatment effects under selection on observables, they allow for the presence of an unobserved covariate that should have been adjusted for in order to estimate the average effect of interest. They explore how strong the correlation between this unobserved covariate and the treatment and the correlation between the unobserved covariate and the potential outcomes would have to be in order the substantially change the estimate for the average effect of interest. A challenge is how to make a case that a particular correlation is substantial or not. \citet{imbens2003} builds on the Rosenbaum and Rubin approach by developing a data-driven way to obtain a set of correlations between the unobserved covariates and treatment and outcome. Specifically he suggests relating the explanatory power of the unobserved covariate to that of the observed covariates in order to calibrate the magnitude of the effects of the unobserved components.

\citet{altonjieldertaber} and \citet{oster} focus on the correlation between the unobserved component in the relation between the outcome and the treatment and observed covariates, and the unobserved component in the relation between the treatment and the observed covariates. In the absence of functional form assumptions this correlation is not identified. 
\citet{altonjieldertaber} and \citet{oster}  therefore explore the sensitivity to fixed values for this correlation, ranging from the case where the correlation is zero (and the treatment is exogenous), to an upper limit, chosen to match the correlation found between the observed covariates in the two regression functions. \citet{oster} takes this further by developing estimators based on this equality. What makes this approach very useful is that for a general set of models it provides the researcher with a systematic way of doing the sensitivity analyses that are routinely, but often in an unsystematic way, done in empirical work.

\subsection{Identification and Sensitivity}
\label{subsection:identification}

\citet{gentzkow2015} 
take a different approach to sensitivity. They propose a method for highlighting what statistical relationships in a dataset are most closely related to parameters of interest.  Intuitively,
the idea is that covariation between particular sets of variables may determine the magnitude of model estimates.  To operationalize this, they investigate in the context of a given model, how the key parameters relate to a set of summary statistics. These summary statistics would typically include easily interpretable functions of the data such as correlations between subsets of variables. Under mild conditions, the joint distribution of the model parameters and the summary statistics should be jointly normal in large samples. If the summary statistics are in fact asymptotically sufficient for the model parameters, the joint distribution of the parameter estimates and the summary statistics will be degenerate. More typically the joint normal distribution will have a covariance matrix with full rank.  \citet{gentzkow2015} discuss how to interpret the covariance matrix in terms of sensitivity of model parameters to model specification. 
\citet{gentzkow2015}  focus on the derivative of the conditional expectation of the model parameters with respect to the summary statistics to assess how important particular summary statistics are for determining the parameters of interest.   More broadly, their approach is related to proposals by
\citet{conley, chetty} in different settings.

\subsection{Supplementary Analyses in Regression Discontinuity Designs}
\label{subsection:suppl_rdd}

One of the most interesting supplementary analyses is the McCrary test in regression discontinuity designs (\citet{mccrary,otsu}). What makes this analysis particularly interesting is the conceptual distance between the primary analysis and the supplementary analysis. The McCrary test assesses whether there is a discontinuity in the density of the forcing variable at the threshold. If the forcing variable is denoted by $X_i$, with density $f_X(\cdot)$, and the threshold $c$, the null hypothesis underlying the McCrary test is
\[ H_0:\ \lim_{x\uparrow c} f_X(x)=\lim_{x\downarrow c} f_X(x),\]
with the alternative hypothesis that there is a discontinuity in the density of the forcing variable at the threshold.
In a conventional analysis, it is unusual that the marginal distribution of a variable that is assumed to be exogenous is of any interest to the researcher: often the entire analysis is conducted conditional on such regressors.

Why is this marginal distribution of interest in this setting? The reason is that the identification strategy underlying regression discontinuity designs relies on the assumption that units just to the left and just to the right of the threshold are comparable. The assumption underling regression discontinuity designs is that it was as good as random on which side of the threshold the units were placed, and implicitly, that there is nothing special about the threshold in that regard. That argument is difficult to reconcile with the finding that there are substantially more units just to the left than just to the right of the threshold. Again, even though such an imbalance is easy to take into account in the estimation, it is the very presence of the imbalance that casts doubt on the entire approach. In many cases where one would find such an imbalance it would suggest that the forcing variable is not a characteristic exogenously assigned to individuals, rather that it is something that is manipulated by someone with knowledge of the importance of the value of the forcing variable for the treatment assignment.

 The classic example is that of an educational regression discontinuity design where the forcing variable is a test score. If the teacher or individual grading the test is aware of the importance of exceeding the threshold, they may assign scores differently than if there were not aware of this. If there was such manipulation of the score, there would likely be a discontinuity in the density of the forcing variable at the threshold: there would be no reason to change the grade for an individual scoring just above the threshold.

Let us return to the Lee election data to illustrate this. For these data the estimated difference in the density at the threshold is 0.10 (with the level of the density around 0.90), with a standard error of 0.08, showing there is little evidence of a discontinuity in the density at the threshold.

\section{Machine Learning and Econometrics}
\label{section:machine}

In recent years there have been substantial advances in flexible methods for analyzing data in computer science and statistics, a literature that is commonly referred to as the
``machine learning'' literature.  These methods have made
only limited inroads into the economics literature, although interest has increased substantially very recently.  There are two broad categories of machine learning, ``supervised'' and
``unsupervised'' learning.  ``Unsupervised learning'' focuses on methods for finding patterns in data, such as groups of similar items.  In the parlance of this review, it focuses
on reducing the dimensionality of covariates  in the absence of outcome data.  Such models have been applied to problems like clustering images or videos, or putting text documents into groups of similar
documents.  Unsupervised learning can be used as a first step in a more complex model.  For example, instead of including as covariates indicator variables for whether a unit (a document) contains each of a very large set of words
in the English language, unsupervised learning can be used to put documents into groups, and then subsequent models could use as covariates indicators for whether a document belongs to one of the groups.  The number of groups might be much smaller than the number of words that appears in all of the documents, and so unsupervised learning is a method to reduce the 
dimensionality of the covariate space.  We do not discuss unsupervised learning further here, beyond simply noting that the method can potentially be quite useful in applications involving text, images,
or other very high-dimensional data, even though they have not had too much use in the economics literature so far (for an exception, see \citet{athey2016aggregators} for an example where unsupervised learning is used to put newspaper articles into topics).  The unsupervised learning literature does have some connections with the statistics literature, for example, for estimating mixture distributions; principal-components analysis is another method that has been used in the social sciences historically, and that falls under 
the umbrella of unsupervised learning.  

``Supervised'' machine learning focuses primarily on prediction problems: given a ``training dataset'' with data on an outcome $Y_i$, which could be discrete or continuous, and some covariates $X_i$, 
the goal is to estimate a model for predicting outcomes in a new dataset (a ``test'' dataset) as a function of $X_i$.   The typical assumption in these methods is that the joint distribution of $X_i$ and $Y_i$ 
is the same in the training and the test data.  Note that this differs from the goal of causal inference in observational studies, where we observe data on outcomes and a treatment variable $W_i$,
and we wish to draw inferences about potential outcomes.  Implicitly, causal inference has the goal of predicting outcomes for a (hypothetical, or counterfactual) test dataset where, for example, the treatment is set to 1 for all
units.  Letting $Y_i^{\obs}=Y_i(W_i)$, by construction, the joint distribution of $W_i$ and $Y_i^{\obs}$ in the training data is different than what it would be in a test dataset where $W_i=1$
for all units.  \citet{kleinberg2015prediction} argue that many important policy problems are fundamentally prediction problems; see also the review article in this volume.  In this review,
we focus primarily on  problems of causal inference, showing how supervised machine learning methods can be used to improve the performance of causal analysis, particularly in cases
with many covariates.  

We also highlight a number of differences in focus between the supervised machine learning literature and the econometrics literature on nonparametric
regression.  A leading difference is that the supervised machine learning literature focuses on how well a prediction model does in minimizing the mean-squared error of prediction in 
an independent test set, often without much attention to the asymptotic properties of the estimator.  The focus on minimizing mean-squared error on a new sample implies that predictions
will make a bias-variance tradeoff; successful methods allow for bias in estimators (for example, by dampening model parameters towards the mean) in order to reduce the variance
of the estimator.  Thus, predictions from machine learning methods are not typically unbiased, and estimators may not be asymptotically normal and centered around the estimand.  Indeed,
the machine learning literature places much less (if any) emphasis on asymptotic normality, and when theoretical properties are analyzed, they often take the forms of worst-case bounds
on risk criteria.

A closely related difference between many (but not all) econometric approaches and supervised machine learning is that many supervised machine learning methods rely on data-driven
model selection, most commonly through cross-validation, to choose ``tuning'' parameters.  Tuning parameters may take the form of a penalty for model complexity, or in the case
of a kernel regression, a bandwidth.  For the supervised learning methods typically the sample is split into two samples, a training sample and a test sample, where
for example the test sample might have $10\%$ of observations.
The training sample is itself partitioned into
a number of subsamples, or cross-validation samples, say $m=1,..,M$, where commonly $M=10$. For each subsample $m=1,\ldots,M$, the cross-validation sample $m$ is set aside. The remainder of the training sample is used for estimation. The estimation results are then used to predict outcomes for the left-out subsample $m$.  The sum of squared residuals for these $M$ subsamples sample are added up. Keeping fixed the partition, the process is repeated for many different values of a tuning parameter.  The final choice of tuning parameter is the one that minimizes the sum of the squared residuals in the cross-validation samples. Cross-validation has been used for kernel regressions within the econometrics literature; in that literature, the convention is often to set $M$ equal to the size of the training sample minus one; that is, researchers often do ``leave-one-out'' cross-validation. In the machine learning literature, the sample sizes are often much larger and estimation may be more complex, so that the computational burden of leave-one-out may be too high.  Thus, the convention is to use $10$ cross-validation samples.  Finally, after the
model is ``tuned'' (that is, the tuning parameter is selected), the researcher re-estimates the model using the chosen tuning parameter and the entire training dataset.  Ultimate
model performance is assessed by calculating the mean-squared error of model predictions (that is, the sum of squared residuals) on the held-out test sample, which was not used at all
for model estimation or tuning. This final step is uncommon in the traditional econometrics literature, where the emphasis is more on efficient estimation and asymptotic properties.

One way to think about cross-validation is that it is tuning the model to best achieve its ultimate goal, which is prediction quality on a new, independent test set.  Since at the time
of estimation, the test set is by definition not available, cross-validation mimics the process of finding a tuning parameter which maximizes goodness of fit on independent samples, since for each
$m$, a model is trained on one sample and evaluated on an independent sample (sample $m$).  The complement of $m$ in the training sample is smaller than the ultimate training sample will
be, but otherwise cross-validation mimics the ultimate exercise.  When the tuning parameter represents model complexity, cross-validation can be thought of as optimizing model complexity 
to balance bias and variance for the estimator.  A complex model will fit very well on the sample used to estimate the model (good in-sample fit), but possibly at the cost of fitting poorly on
a new sample.  For example, a linear regression with as many parameters as observations fits perfectly in-sample, but may do very poorly on a new sample, due to what is referred to as
``over-fitting.''

The fact that model performance (in the sense of predictive accuracy on a test set) can be directly measured makes it possible to meaningfully compare predictive models, even when their
asymptotic properties are not understood.  It is perhaps not surprising that enormous progress has been made in the machine learning literature in terms of developing models that 
do well (according to the stated criteria) in real-world datasets.
Here, we  briefly review some of the supervised machine learning methods that are most popular and also most useful for causal inference, and relate them to methods traditionally used in the economics and econometrics literatures.  We then describe some of the recent literature combining machine learning and econometrics for causal inference.

\subsection{Prediction Problems}
\label{subsection:prediction}

The first problem we discuss is that of nonparametric estimation of regression functions. The setting is one where we have observation for a number of units on an outcome, denoted by $Y_i$ for unit $i$, and a vector of features, covariates, exogenous variables, regressors or predictor variables, denoted by $X_i$. The dimension of $X_i$ may be large, both relative to the number of units and in absolute terms. 
The target is the conditional expectation
\[ g(x)=\mme[Y_i|X_i=x].\]
For this setting, the traditional methods in econometrics are based on kernel regression or nearest neighbor methods (\citet{hardle, wasserman}). In ``K-nearest-neighbor'' or KNN methods, $\hat{g}(x)$ is
the sample average of the $K$ nearest observations to $x$ in Euclidean distance.  $K$ is a tuning parameter; when applied in the supervised machine learning
literature, $K$ might be chosen through cross-validation to minimize mean-squared error on independent test sets.  In economics, where bias-reduction is often paramount, it is more common
to use a small number for $K$.    Kernel regression is similar, but a weighting function is used to weight observations nearby to $x$ more heavily than those far away.  Formally, the kernel regression the estimator $\hat g(x)$ has the form
\[  \hat g(x)=\sum_{i=1}^N Y_i\cdot K\left(\frac{X_i-x}{h}\right) \Bigl/  
\sum_{i=1}^N K\left(\frac{X_i-x}{h}\right) ,\]
for some kernel function $K(\cdot)$, sometimes  a normal kernel $K(x)=\exp(-x^2/2)$, or  bounded kernel such as the uniform kernel $K(x)=\mathbf{1}_{|x|\leq 1}$.
The properties of such kernel estimators are well established, and known to be poor when the dimension of $X_i$ is high. To see why,
note that with many covariates, the nearest observations across a large number of dimensions may not be particularly close in any given dimension.  

Other alternatives for nonparametric regression include series regression where $g(x)$ is approximated by the sum of a set of basis functions, $g(x)=\sum_{k=0}^K \beta_k\cdot h_k(x)$, for example polynomial basis functions, $h_k(x)=x^k$ (although the polynomial basis is rarely an attractive choice in practice). These methods do have well established properties (\citet{neweymcfadden}), including asymptotic normality, but they do not work well in high-dimensional cases.  


\subsubsection{Penalized Regression}
\label{subsubsection:LASSO}

One of the most important methods in the supervised machine learning literature is the class of penalized regression models, where one of the most popular members of this class is LASSO (Least Absolute Shrinkage and Selection Operator, \citet{tibshirani1996regression, hastie2009elements, hastie2015statistical}). This estimator imposes a linear model for outcomes as a function of covariates and attempts to minimize an objective that includes the sum of square residuals
as in ordinary least squares, but also adds on an additional term penalizing the magnitude of regression parameters.  Formally, the objective function for these penalized regression models, after demeaning the covariates and outcome, and standardizing the variance of the covariates, can be written as
\begin{equation} 
\min_{\beta_1,\ldots,\beta_K} \sum_{i=1}^N
\Bigl( Y_i-\sum_{k=1}^K \beta_k\cdot X_{ik}\Bigr)^2
+\lambda\cdot 
\left\|\beta\right\|,\label{eqn:LASSO}
\end{equation}
where $\left\|\cdot\right\|$ is a general norm.
 The standard practice is to select the tuning parameter $\lambda$ through cross-validation.  To interpret this, note that if we take $\lambda=0$, we are back in the least squares world, and obtain the ordinary least squares estimator. However, the ordinary least squares estimator is not unique if there are more regressors than units, $K>N$. Positive values for $\lambda$ regularize this problem, so that the solution to the LASSO minimization problem is well defined even if $K>N$. With a positive value for $\lambda$, there are a number of interesting choices for the norm. A key feature is that for some choices of the norm, the algorithm leads to some of the $\beta_k$ to be exactly zero, leading to a sparse model. For example,  the $L_0$ norm  $\|\beta\|=\sum_{k=1}^K \mathbf{1}_{\beta_k\neq 0}$ leads to optimal subset selection: the estimator selects some of the $\beta_k$ to be exactly zero, and estimates the remainder by ordinary least squares. Another interesting choice is the $L_2$ norm, $\|\beta\|=\sum_{k=1}^K \beta_k^2$ , which leads to ridge regression: all $\beta_k$ are shrunk smoothly towards zero, but none are set equal to zero. In that case there is a very close connection to Bayesian estimation. If we specify the prior distribution on the $\beta_k$ to be Gaussian centered at zero, with variance equal to $\lambda$, the estimator for $\beta$ is equal to the posterior mean.
Perhaps the most important case is $\|\beta\|=\sum_{k=1}^K |\beta_k|$. In that case some of the $\beta_k$ will be estimated to be exactly equal to zero, and the remainder will be shrunk towards zero. This is the LASSO (\citet{tibshirani1996regression, hastie2009elements, hastie2015statistical}). The value of the tuning parameter $\lambda$ is typically choosen by cross-validation. 

Consider the choice between LASSO and ridge regression. From a Bayesian perspective, both can be interpreted as putting independent prior distributions on all the $\beta_k$, with in one case the prior distributions being normal and in the other case the prior distributions being Laplace. There appears to be little reason to favor one rather than the other conceptually. 
 \citet{tibshirani1996regression} in the original LASSO paper discusses scenarios where LASSO performs better (many of the $\beta_k$ equal or very close to zero, and a few that are large), and some where ridge regression performs better (all $\beta_k$ small, but not equal to zero).
The more important difference is that LASSO leads to a sparse model. This can make it easier to interpret and discuss the estimated model, even if it does not perform any better in terms of prediction than ridge regression. Researchers should ask themselves whether the sparsity is important in their actual application. If the model is simply used for prediction, this feature of LASSO may not be of intrinsic importance.  Computationally effective algorithms have been developed that allow for the calculation of the LASSO estimates in large samples with many regressors. 

One important extension that has become popular is to combine the ridge penalty term that is proportional to 
$(\sum_{k=1}^K |\beta_k|^2)$ with the LASSO penalty term that is proportional to
$\sum_{k=1}^K |\beta_k|$ in what is called an elastic net (\citet{hastie2009elements, hastie2015statistical}).  There are also many extensions of the basic LASSO methods, allowing for nonlinear regression (e.g., logistic regression models) as well as selection of groups of parameters, see 
  \citet{hastie2009elements, hastie2015statistical}.

Stepping back from the details of the choice of norm for penalized regression, one might consider why the penalty term is needed at all outside the case where there are more
covariates than observations.  For smaller values of $K$, we can return to the question of what the goal is of the estimation procedure.  Ordinary least squares is unbiased; it
also minimizes the sum of squared residuals for a given sample of data.  That is, it focuses on in-sample goodness-of-fit.  One can think of the term involving the penalty in (\ref{eqn:LASSO})
as taking into account the ``over-fitting'' error, which corresponds to the expected difference between in-sample goodness of fit and out-of-sample goodness of fit.  Once
covariates are normalized, the magnitude of $\beta$ is roughly proportional to the potential of the model to over-fit.  Although the gap between in-sample and out-of-sample fit
is by definition unobserved at the time the model is estimated, when $\lambda$ is chosen by cross-validation, its value is chosen to balance in-sample and out-of-sample prediction in a way
that minimizes mean-squared error on an independent dataset.

Unlike many supervised machine learning methods, there is a large literature on the formal asymptotic properties of the LASSO; this may make the LASSO more attractive as an empirical 
method in economics. Under some conditions standard least squares confidence intervals based ingoring the variable selection feature of the LASSO are valid. The key condition is that the true value for many of the regressors is in fact exactly equal to zero, with the number of non-zero parameter values increasing very slowly with the sample size. 
See
  \citet{hastie2009elements, hastie2015statistical}. 
This condition is of course unlikely to hold exactly in applications.
LASSO is doing data-driven model selection, and ignoring the model selection for inference as suggested by the theorems based on  these sparsity assumptions may lead to substantial under-coverage for confidence intervals in practice.  In addition, it is important to recognize that regularized regression models reward parsimony: if there are several correlated variables, LASSO
will prefer to put more weight on one and drop the others.  Thus, individual coefficients should be interpreted with caution in moderate sample sizes or when sparsity is not known to hold.  

\subsubsection{Regression Trees}

Another important class of methods for prediction that is only now beginning to make inroads into the economics literature is regression trees and its generalizations. The classic reference for regression trees is \citet{breiman1984classification}. Given sample with $N$ units and a set of regressors $X_i$, the idea is to sequentially partition the covariate space into subspaces in a way that reduces the sum of squared residuals as much as possible. Suppose, for example, that we have two covariates $X_{i1}$ and $X_{i2}$. 
Initially the sum of squared residuals is $\sum_{i=1}^N (Y_i-\overline{Y})^2$.
We can split the sample by $X_{i1}<c$ versus $X_{i1}\geq c$, or we can split it by $X_{i2}<c$ versus $X_{i2}\geq c$. We look for the split (either splitting by $X_{i1}$ or by $X_{i2}$, and the choice of $c$) that minimizes the sum of squared residuals.  After the first split we look at the two subsets (the two leaves of the tree), and we consider the next split for each of the two subsets. At each stage there will be a split (typically unique) that reduces the sum of squared residuals the most.  In the simplest version of a regression tree we would stop once the reduction in the sum of squared residuals is below some threshold. We can think of this as adding a penalty term to the sum of squared residuals that is proportional to the number of leaves.
A more sophisticated version of the regression trees first builds (grows) a large tree, and then prunes leaves that have little impact on the sum of squared residuals. This avoids the problem that a simple regression tree may miss splits that would lead to subsequent profitable splits if the initial split did not improve the sum of squared residuals sufficiently.
In both cases a key tuning parameter is the penalty term on the number of leaves. The standard approach in the literature is to choose that through crossvalidation, similar to that discussed in the LASSO section. 

There is relatively little asymptotic theory on the properties of regression trees. Even establishing consistency for the simple version of the regression tree, let alone inferential results that would allow for the construction of confidence intervals is not straightforward. A key problem in establishing such properties is that the estimated regression function is a non-smooth step function.

We can compare regression trees to common practices in applied work of capturing nonlinearities in a variable by discretizing the variable, for example, by dividing it into deciles.  The regression tree uses the data to determine the appropriate ``buckets'' for
discretization, thus potentially capturing the underlying nonlinearities with a more parsimonious form.  On the other hand, the
regression tree has difficulty when the underlying functional form is truly linear.

Regression trees are generally dominated by other, more continuous models when the only goal is prediction.  Regression trees
are used in practice due to their simplicity and interpretability.  Within a partition, the prediction from a regression tree is simply
a sample mean.  Simply by inspecting the tree (that is, describing the partition), it is straightforward to understand why a particular
observation received the prediction it did.  

\subsubsection{Random Forests}

Random forests are one of the most popular supervised machine learning methods, known for their reliable ``out-of-the-box''
performance that does not require a lot of model tuning.  They perform well in prediction contests; for example, in a recent economics paper (\citet{glaeser2016predictive})
on crowd-sourcing predictive algorithms for city governments through contests, the winning algorithm was a random forest.  

One way to think about random forests is that they are are an example of ``model averaging.'' The prediction of a random forest
is constructed as the average of hundreds or thousands of distinct regression trees.  The regression trees differ from one another
for several reasons.  First, each tree is constructed on a distinct training sample, where the samples are selected by either
bootstrapping or subsampling.  Second, at each potential split in constructing the tree, the algorithm considers a random subset
of covariates as potential variables for splitting.  Finally, each individual tree is not pruned, but typically is ``fully grown'' up to
some minimum leaf size.  By averaging distinct predictive trees, the discontinuities of regression trees are smoothed out, and each
unit receives a fully personalized prediction.

Although the details of the construction of random forests are complex and look quite different than standard econometric methods,
\citep{wagerathey} argue that random forests are closely related to other non-parameteric methods such as k-nearest-neighbor
algorithms and kernel regression.  The prediction for each point is a weighted average of nearby points, since each underlying
regression tree makes a prediction based on a simple average of nearby points, equally weighted.  The main conceptual difference
between random forests and the simplest versions of nearest neighbor and kernel algorithms is that there is a data-driven approach
to select which covariates are important for determining what data points are ``nearby'' a given point.  However, using
the data to select the model also comes at a cost, in that the predictions of the random forest are asymptotically bias-dominated.

Recently, \cite{wagerathey} develop a modification of the random forest where the predictions are asymptotically normal and
centered around the true conditional expectation function, and also propose a consistent estimator for the asymptotic variance, so
that confidence intervals can be constructed.  The most important deviation from the standard random forest is that two
subsamples are used to construct each regression tree, one to construct the partition of the covariate space, and a second
to estimate the sample mean in each leaf.  This sample splitting approach ensures that the estimates from each component tree
in the forest are unbiased, so that the predictions of the forest are no longer asymptotically bias-dominated.  Although asymptotic
normality may not be crucial for pure prediction problems, when the random forest is used as a component of estimation of
causal effects, such properties play a more important role, as we show below.

\subsubsection{Boosting}

A general way to improve simple machine learning methods is boosting. We discuss this in the context of regression trees, but its application is not limited to such settings. Consider a very simple algorithm for estimating a conditional mean, say a tree with only two leaves. That is, we only split the sample once, irrespective of the number of units or the number of features. This is unlikely to lead to a very good predictor. The idea behind boosting is to repeatedly apply this naive method. After the first application we calculate the residuals. We then apply the same method to the residuals instead of the original outcomes. That is, we again look for the sample split that leads to the biggest reduction in the sum of squared residuals. We can repeat this many times, each time applying the simple single split regression tree to the residuals from the previous stage.

If we apply this simple learner many times, we can approximate the regression function in a fairly flexible way. However, this does not lead to an accurate approximation for all regression functions. By limiting ourselves to a naive learner that is a single split regression tree we can only approximate additive regression functions, where the regression function is the sum of functions of one of the regressors at a time. If we want to allow for interactions between pairs of the basic regressors we need to start with a simple learner that allows for two splits rather than one.

\subsubsection{Super Learners and Ensemble methods}

One theme in the supervised machine learning literature is that model averaging often performs very well; many contests such as
those held by Kaggle are won by algorithms that average many models.  Random forests use a type of model averaging, but all 
of the models that are averaged are in the same family.  In practice, performance can be better when many different types of models
are averaged.  The idea of Super Learners in \citet{van2007super} is to use model performance to construct weights, so that better performing models receive more weight in the averaging.

\subsection{Machine Learning Methods for Average Causal Effects}
\label{subsection:machine_average}

There is a large literature on estimating  treatment effects in settings with selection on observables, or unconfoundedness. This literature has largely focused on the case with a fixed and modest number of covariates. In practice, in order to make the critical assumptions more plausible, the number of pretreatment variables may be substantial.  In recent years, researchers have introduced machine learning methods into this literature to account for the presence of many covariates. In many cases, the newly proposed estimators closely mimic estimators developed in the literature with a fixed number of covariates.  From a conceptual perspective, being able to flexibly control for a large number of covariates may make an estimation strategy much more convincing, particularly if the identification assumptions are 
only plausible once a large number of confounding variables have been controlled for.

\subsubsection{Propensity Score Methods}

One strand of the literature has focused on estimators that directly involve the propensity score, either through weighting or matching. Such methods had been shown in the fixed number of covariates case to lead to semiparametrically efficient estimators for the average treatment effect, e.g., \citet{hahn1998role, hirr}. The specific implementations in those papers, relying on kernel or series estimation of the propensity score,  
would be unlikely to work in settings with many covariates.

In order to deal with many covariates, researchers have proposed estimating the propensity score using random forests, boosting, or LASSO, and then use weights based on those estimates following the usual approaches from the existing literature (e.g., \citet{mccaffrey2004propensity, wyss}). One concern with these methods is that even in settings with few covariates the weighting and propensity matching methods have been found to be sensitive to the implementation of the propensity score estimation. Minor changes in the specification, e.g. using logit models versus probit models, can change the weights substantially for units with propensity score values close to zero or one, and thus lead to estimators that lack robustness. Although the modern nonparametric methods may improve the robustness somewhat compared to previous methods, the variability in the weights is not likely to  improve with the presence of many covariates.  Thus, procedures such as ``trimming'' the data to eliminate extreme values of the estimated propensity score (thus changing the estimand as in \citep{crump2009dealing}) remain important.

\subsubsection{Regularized Regression Methods}

\citet{belloni2014inference, belloni2014jep, belloni2013program} focus on regression estimators for average treatment effects. For ease of exposition, suppose one is interested in the average effect for the treated, and so the problem is to estimate $\mathbb{E}[Y(0)|W_i=1]$. Under unconfoundedness this is equal to $\mathbb{E}[\mathbb{E}[Y_i^\obs|W_i=0,X_i]|W_i=1]$. Suppose we model $\mathbb{E}[Y_i^\obs|X_i=x,W_i=0]$ as $x'\beta_c$. 
\citet{belloni2014inference} point out that estimating $\beta_c$ using lasso leads to estimators for average treatment effects with poor properties. Their insight is that the objective function for LASSO (which is purely based on predicting outcomes) leads the LASSO to select covariates that are highly correlated with the outcome; but the objective fails to prioritize covariates that are highly correlated with the treatment but only weakly correlated with outcomes.  Such variables are potential confounders for the average treatment effect,
and omitting them leads to bias, even if they are not very important for predicting unit-level outcomes. This highlights a general issue
with interpreting individual coefficients in a LASSO: because the LASSO objective focuses on prediction of outcomes rather than unbiased
estimation, individual parameter estimates should be interpreted with caution.  LASSO penalizes the inclusion of covariates, and some
will be omitted in general; LASSO will favor a more parsimonious functional form, where if two covariates are correlated, only one will be included, and its parameter estimate will reflect the effects of both the included and omitted variables.  Thus, in general LASSO 
coefficients should not be given a causal interpretation.

\citet{belloni2013program} propose a modification of the LASSO that addresses these concerns and restores the ability of LASSO
to produce valid causal estimates.  They propose a double selection procedure, where they use LASSO first to select covariates that are correlated with the outcome, and then again to select covariates that are correlated with the treatment. In a final ordinary least squares regression they include the union of the two sets of covariates, greatly improving the properties of the estimators for the average treatment effect.  This approach accounts for omitted variable bias that would otherwise appear in a standard LASSO.  \citet{belloni2014jep} illustrate the magnitude of the bias that can occur in real-world datasets from failing to account for this issue.  More broadly, these papers highlight the distinction between predictive modeling and estimation of causal effects.

\subsubsection{Balancing and Regression}

An alternative line of research has focused on finding weights that directly balance covariates or functions of the covariates between
treatment and control groups, so that once the data has been re-weighted, it mimics more closely a randomized experiment. In the earlier literature with few covariates, this approach has been developed in
\citet{hainmueller, graham1, graham2}. More recently these ideas have also been applied to the many covariates case in
\citet{zubizarreta2015stable, imai2014covariate}.
\citet{atheyimbenswager} develop an estimator that combines the balancing with regression adjustment, in the spirit of the double robust estimators proposed by \citet{robins1, robins2, schafer}.  The idea is that, in order to predict the counterfactual outcomes
that the treatment group would have had in the absence of the treatment, it is necessary to extrapolate from control observations.
By rebalancing the data, the amount of extrapolation required to account for differences between the two groups is reduced.
To capture remaining differences, regularized regression can be used to model outcomes in the absence of the treatment.

The general form of the \citet{atheyimbenswager} estimator for the expected control outcome for the treated, that is, $\mu_c=\mathbb{E}[Y_i(0)|W_i=1]=\mathbb{E}[Y_i|X_i=x,W_i=0]$,  is
\[
\hmucc =  \hatx \cdot \hbetacc + \sum_{{i : W_i = 0}} \gamma_i \, \left(Y_i^\obs - X_i \cdot \hbetacc \right). 
\]
They suggest estimating $\hbetacc $ using LASSO or elastic net, in a regression of $Y^\obs_i$ on $X_i$ using the control units. They suggest choosing the weights $\gamma_i$ as the solution to
\[
\gamma = \arg\min_{\gamma} {(1 - \zeta) \left\|\gamma\right\|_2^2 + \zeta \left\|\hatx - \bxcc^\top \gamma\right\|_\infty^2 \text{ subject to } \sum \gamma_i = 1,\ \gamma_i\geq 0}.
\]
This objective function balances the bias coming from imbalance between the covariates in the treated subsample and the weighted control subsample and the variance from having excessively variable weights.
They suggest using $\zeta=1/2$.  Unlike methods that rely on directly estimating the treatment assignment process (e.g. the propensity score), the method controls bias even when the process determining treatment assignment cannot be represented with a sparse model.

\subsection{Heterogenous Causal Effects}
\label{subsection:machine_hetero}
A different problem is that of estimating the average effects of the treatment for each value of the features, that is, the conditional
average treatment effect (CATE) $\tau(x)=\mathbb{E}[Y_i(1)-Y_i(0)|X_i=x]$. This problem is highly relevant as a step towards assigning units to optimal treatments. If all costs and benefits of the treatment are incorporated in the measured outcomes, understanding the set of covariates where CATE is positive all that matters for determining treatment assignment; in contrast, if the policy might be applied in different settings with
additional costs or benefits that might be different than those in the training data, or if the analyst wants to also gain insight about treatment effect heterogeneity,

The concern is that searching over many covariates and subsets of the covariate space may lead to spurious findings of treatment effect differences.  Indeed, in medicine (e.g. for clinical trials), pre-analysis plans must be registered in advance
to avoid the problem that researchers will be tempted to search for heterogeneity, and may instead end up with spurious findings.
This problem is more severe when there are many covariates.


\subsubsection{Multiple Hypothesis Testing}

 One approach to this problem is to exhaustively search for treatment effect
heterogeneity and then correct for issues of multiple testing.  By multiple testing, we mean the problem that arises when a researcher
considers a large number of statistical hypotheses, but analyzes them as if only one had been considered.  This can lead to ``false
discovery,'' since across many hypothesis tests, we expect some to be rejected even if the null hypothesis is true.  

To address this problem, \citet{list2016multiple} propose to discretize
each covariate, and then loop through the covariates, testing whether the treatment effect is different when the covariate is
low versus high.  Since the number of covariates may be large, standard approaches to correcting for multiple testing may
severely limit the power of a (corrected) test to find heterogeneity.  \citet{list2016multiple} propose an approach based on
bootstrapping that accounts for correlation among test statistics; this approach can provide substantial improvements over
standard multiple testing approaches when the covariates are highly correlated, since dividing the sample according to each of two
highly correlated covariates results in substantially the same division of the data.  

A drawback of this approach is that the researcher must specify in advance all of the hypotheses to be tested; alternative
ways to discretize covariates, and flexible interactions among covariates, may not be possible to fully explore.  A different
approach is to adapt machine learning methods to discover particular forms of heterogeneity, as we discuss
in the next section.

\subsubsection{Subgroup Analysis}

In some settings, it is useful to identify subgroups that have different treatment effects.  One example is where eligibility for
a government program is determined according to various criteria that can be represented in a decision tree, or when a doctor
uses a decision tree to determine whether to prescribe a drug to a patient.  Another example is when an algorithm uses
a simple lookup table to determine which type of user interface, offer, email solicitation, or ranking of search results to provide to a user.
Subgroup analysis has long been used in medical studies (\citep{foster2011subgroup}), but it is often subject to criticism due to concerns of multiple testing (\citep{assmann2000subgroup}).

\citet{athey2016recursive} develops a method that they call ``causal trees.'' The method is based on regression trees, and its goal is to identify a partition of the covariate space into 
subgroups based on treatment effect heterogeneity.  The output of the method is a treatment effect and a confidence interval for each subgroup.  The approach
differs from standard regression trees in several ways.  First, it uses a different criterion for building the tree: rather than
focusing on improvements in mean-squared error of the prediction of outcomes, it focuses on mean-squared error of treatment effects.
Second, the method relies on ``sample splitting'' to ensure that confidence intervals have nominal coverage, even when the
number of covariates is large.  In particular, half the sample is used to determine the optimal partition of the covariates space
(the tree structure), while the other half is used to estimate treatment effects within the leaves.

\citet{athey2016recursive} highlight the fact that the criteria used for tree construction and cross-validation should differ when
the goal is to estimate treatment effect heterogeneity rather than heterogeneity in outcomes; the factors that affect the level 
of outcomes might be quite different from those that affect treatment effects.  To operationalize this, the criteria used for sample splitting and cross-validation must confront two problems.  First, unlike individual outcomes,
the treatment effect is not observed for any individual in the dataset.  Thus, it is not possible to directly calculate a sample
average of the mean-squared error of treatment effects, as this criterion is infeasible:
\begin{equation}
-\frac{1}{N} \sum_{i=1}^N
\Bigl( \tau_i - \hat\tau(X_i)\Bigr)^2.
\end{equation}
However, the approach exploits the fact that the regression tree makes the same prediction within each leaf.  Thus, the
estimator $\hat\tau$ is constant within a leaf, and so the infeasible mean-squared error criterion can be estimated, since
it depends only on averages of $\tau_i$ within leaves.  
The second issue is that the criteria are adapted to anticipate the fact that the model will be re-estimated with an independent
data set.  The modified criterion rewards a partition that creates differentiation in estimated treatment effects, but penalizes
a partition where the estimated treatment effects have high variance, for example due to small sample size.

Although the sample-splitting approach may seem extreme--ultimately only half the data is used for estimating treatment
effects--it has several advantages.  One is that the confidence intervals are valid no matter how many covariates are used in estimation.
The second is that the researcher is free to estimate a more complex model in the second part of the data--the partition can be used
to create covariates and motivate interactions in a more complex model, for example if the researcher wishes to include fixed effects in the model, or model different types of correlation in the error structure.  

Other related approaches include \citet{su2009subgroup} and \citet{zeileis2008model}, who propose statistical tests as criteria
in constructing partitions.  Neither of these approaches address the issue of constructing valid confidence intervals using the
results of the partitions, but \citet{athey2016recursive} combines their approaches with sample splitting in order to obtain valid
confidence intervals on treatment effects.  The approach of \citet{zeileis2008model} is more general than the problem of estimating treatment effect heterogeneity: this paper proposes estimating a potentially rich model within each leaf of the tree, and the criterion for
splitting a leaf of the tree is a statistical test based on whether the split improves goodness of fit of the model.

\subsubsection{Personalized Treatment Effects}

\citet{wagerathey} propose a method for estimating heterogeneous treatment effects based on random forests.  Rather than
rely on the standard random forest model, which focuses on prediction, \citet{wagerathey} build random forests where each
component tree is a causal tree \citep{athey2016recursive}.  Relative to a causal tree, which identifies a partition and estimates
treatment effects within each element of the partition, the causal forest leads to smooth estimates of $\tau(x)$.  This type of method
is more similar to a kernel regression, nearest-neighbor matching, or other fully non-parametric methods, in that a distinct prediction 
is provided for each value of $x$.  Building on their work for prediction-based random forests, \citet{wagerathey} show that 
the predictions from causal forests are asymptotically normal and centered on the true CATE for each $x$, since causal trees make use of sample splitting.  They also propose an estimator for the variance, so that confidence intervals can be obtained.  Relative to existing methods from econometrics, the random 
forest has been widely documented to perform well (for prediction problems) in a variety of settings with many covariates; and
a particular advantage over methods such as nearest neighbor matching is that the random forest is resilient in the face of many
covariates that have little effect.  These covariates are simply not selected for splitting when determining the partition.  In contrast, nearest neighbor matching deteriorates quickly with additional irrelevant covariates.

An alternative approach, closely related, is based on Bayesian Additive Regression Trees (BART) \citep{chipman2010bart}.  \citet{hill2011bayesian} and \citet{green2012modeling} apply these methods to estimate heterogeneous treatment effects.  BART
is essentially a Bayesian version of random forests.  Large sample properties of this method are unknown, but it appears to have
good empirical performance in applications.

Another approach is based on the LASSO \citep{imai2013estimating}.  This approach estimates a LASSO model with the treatment indicator
interacted with covariates, and uses LASSO as a variable selection algorithm for determining which covariates are most important.
In order for confidence intervals to be valid, the true model must be assumed to be sparse.  It may be prudent in a particular datset
to perform some supplementary analysis to verify that the method is not over-fitting; for example, one could test the approach by using only half of the data to estimate the LASSO, and then comparing the results to an ordinary least squares regression with the variables selected by LASSO in the other half of the data.  If the results are inconsistent, it could simply indicate that using half the data is not good enough; but it also might indicate that sample splitting is warranted to protect against over-fitting or other sources of bias that arise
when data-driven model selection is used.

A natural application of personalized treatment effect estimation is to estimating optimal policy functions.  A literature in machine learning considers this 
problem (\citep{beygelzimer2009offset}; \citep{langford2011doubly}); some open questions include the ability to obtain confidence intervals on differences between policies obtained from these methods.  The machine learning literature tends to focus more on
worst-case risk analysis rather than confidence intervals.   

\subsection{Machine Learning Methods with Instrumental Variables}
\label{subsection:machine_iv}

Another setting where high-dimensional predictive methods can be useful is in settings with instrumental variables. The first stage in instrumental variables is typically purely a predictive exercise, where the conditional expectation of the endogenous variables is estimated using all the exogenous variables and excluded instruments.
If there are many instruments, and these can arise from a few instruments interacted with indicators for subpopulations, or from other flexible transformations of the basic instrument, standard methods are known to have poor properties (\citet{stock1997}).
Alternative methods have focused on asymptotics based on many instruments (\citet{bekker1994}), or hierarchical Bayes or random effects methods (\citet{chamberlain}).  It is possible to interpret the latter approach as instituting a form of ``shrinkage'' similar to ridge.

\citet{belloni2013program} develop LASSO methods to estimate the first (as well as second) stage in such settings, providing conditions
under which valid confidence intervals can be obtained.

In a different setting \citet{bakshy} study the use of instrumental variables in network settings. Encouragement to take particular actions that affects friends an individual is connnected to is randomized at the individual level. These then generate many instruments that each only weakly affect a particular individual.

\section{Conclusion} \label{section:conclusion}

This review has covered selected topics in the area of causality and policy evaluation.  We have attempted to highlight recently developed approaches for estimating the impact
of policies.  Relative to the previous literature, we have tried to place more emphasis on supplementary analyses that help the analyst assess the credibility of estimation and identification
strategies.  We further review recent developments in the use of machine learning for causal inference; although in some cases, new estimation methods have been proposed, we also
believe that the use of machine learning can help buttress the credibility of policy evaluation, since in many cases it is important to flexibly control for a large number of covariates as
part of an estimation strategy for drawing causal inferences from observational data.  We believe that in the coming years, this literature will develop further, helping researchers avoid
unnecessary functional form and other modeling assumptions, and increasing the credibility of policy analysis. 

\bibliographystyle{plainnat}

\bibliography{references}

\end{document}